\documentstyle[aps,multicol,pra,tighten,eqsecnum,graphicx]{revtex}
\begin{document}

\draft

\title{Adiabatic Elimination in Compound Quantum Systems with Feedback}
\author{P. Warszawski${}^{1,2}$ and H.M. Wiseman${}^{1,2}$}
\address{${}^{1}$ School of Science, Griffith University, Nathan, Brisbane,
Queensland 4111 Australia\\
${}^{2}$Department of Physics, The University of Queensland,
St Lucia, Brisbane, Queensland 4072 Australia. }
\maketitle

\begin{abstract}
Feedback in compound quantum systems is effected by using the output
from one sub-system (``the system'')  to control the
evolution of a second
sub-system (``the ancilla'') which is reversibly coupled to the system.
In the limit where the ancilla responds to
fluctuations on a much shorter time scale than does the system, we show
that it can be
adiabatically eliminated, yielding a master equation for
the system alone.  This is very significant as it
decreases the necessary basis size for numerical
simulation and allows the effect of the ancilla to be understood more
easily.  We consider two types of ancilla: a two-level ancilla (e.g.
a two-level atom) and an infinite-level ancilla (e.g.  an optical
mode).  For each, we consider two forms of feedback:
coherent (for which a quantum
mechanical description of the feedback loop is required) and incoherent
(for which a classical description is sufficient).
We test the master equations we obtain using
numerical simulation of the full dynamics of the compound system.
For the system (a parametric oscillator) and feedback
(intensity-dependent detuning) we choose, good agreement is found in
the limit of heavy damping of the ancilla. We discuss the relation of
our work to previous work on feedback in compound quantum systems,
and also to previous work on adiabatic elimination in general.
\end{abstract}

\newcommand{\beq}{\begin{equation}}
\newcommand{\eeq}{\end{equation}}
\newcommand{\bqa}{\begin{eqnarray}}
\newcommand{\eqa}{\end{eqnarray}}
\newcommand{\nn}{\nonumber}
\newcommand{\nl}[1]{\nn \\ && {#1}\,}
\newcommand{\erf}[1]{Eq.~(\ref{#1})}
\newcommand{\erfs}[2]{Eqs.~(\ref{#1})--(\ref{#2})}
\newcommand{\dg}{^\dagger}
\newcommand{\rt}[1]{\sqrt{#1}\,}
\newcommand{\smallfrac}[2]{\mbox{$\frac{#1}{#2}$}}
\newcommand{\half}{\smallfrac{1}{2}}
\newcommand{\bra}[1]{\langle{#1}|}
\newcommand{\ket}[1]{|{#1}\rangle}
\newcommand{\ip}[2]{\langle{#1}|{#2}\rangle}
\newcommand{\sch}{Schr\"odinger }
\newcommand{\schs}{Schr\"odinger's }
\newcommand{\hei}{Heisenberg }
\newcommand{\heis}{Heisenberg's }
\newcommand{\bl}{{\bigl(}}
\newcommand{\br}{{\bigr)}}
\newcommand{\ito}{It\^o }
\newcommand{\str}{Stratonovich }
\newcommand{\dbd}[1]{\frac{\partial}{\partial {#1}}}
\newcommand{\sq}[1]{\left[ {#1} \right]}
\newcommand{\cu}[1]{\left\{ {#1} \right\}}
\newcommand{\ro}[1]{\left( {#1} \right)}
\newcommand{\an}[1]{\left\langle{#1}\right\rangle}
\newcommand{\implies}{\Longrightarrow}
\newcommand{\ve}{\varepsilon}

\begin{multicols}{2}

\section{Introduction}
The quantum theory of continuous Markovian feedback is now well
understood \cite{WisMil93b,WisMil94a,Wis94a,WisMil94b}. Continuous
feedback arises in a situation where a system continuously interacts with
its environment, and the environment is deliberately engineered such
that the influence of the system on the environment acts back on the
system at a later time. This can be described as a Markovian process
when (a) the natural coupling of the system to the environment is
approximately Markovian, and (b) the effective time-delay
in the feedback process is negligible compared to any relevant time
scale of the system. If the Markovian approximation is appropriate,
this leads to the great simplification that the system evolution may
be described by a master equation of the Lindblad form \cite{lind}.

It is possible to divide quantum
feedback into two categories, which we may call coherent and
incoherent, following Lloyd \cite{Llo98} (but without being limited by
his definitions).
In the latter case of incoherent feedback, it is not necessary to use
a quantum description of the entire feedback loop. Rather, at some
point, it is permissible to change from a quantum to a classical
description by invoking a measurement step. In a quantum optical
context, this corresponds to electro-optical feedback \cite{WisMil94b}
where a
photocurrent derived from detecting the light radiated by the system
is used to control electro-optical devices which change the behavior
of the system. In the former case of coherent feedback, a quantum
description of the entire feedback loop is necessary. In a quantum
optical context this corresponds to all-optical feedback
\cite{WisMil94b} in which the light radiated by the system is
reflected so that in interacts with the system again, perhaps via some
other system.

Continuous quantum feedback may be non-Markovian for a number of
reasons. The coupling to the environment may be non-Markovian. The
time delay in the feedback loop may be non-negligible. The feedback
may act via a second system, the ancilla. In this paper we are
concerned with the last possibility. This is of interest because it
arises very naturally in quantum optics in both all-optical
\cite{WisMil94b} and electro-optical \cite{SloMil94} contexts.
In principle, this sort of
feedback can be described as a Markovian process in the larger
state space of the system plus ancilla. In practice, this procedure is
often not useful, because of the critical word {\em larger} in the
previous sentence. If the required basis size of the
system and ancilla are $N$ and $M$ respectively, then the Liouvillian
for the compound system has of order $N^{4}M^{4}$ elements. Clearly for
$M$ large, this is much larger than a Liouvillian for the system alone.

Consequently it would be an advantage to obtain a master equation for the
system alone, without the ancilla. This is possible if the ancilla
can be adiabatically eliminated. That is, if the ancilla has a decay
rate much faster than any relevant system rate, so that it is
always in a steady state determined by the system state. It is the
purpose of this paper to determine numerically
the conditions under which this is
possible, and to derive the resultant master equations under those
conditions, for a variety
of general feedback systems.

Previous work in this area has left the situation somewhat confused.
Wiseman and Milburn \cite{WisMil94b}
considered all-optical feedback via an ancilla
system, and adiabatically eliminated the ancilla. This was shown to
be equivalent to electro-optical feedback for quadrature feedback.
However, for intensity feedback it was the same only to second order in
the feedback strength. Moreover, the master equation derived (to
second order) was not of the Lindblad form.

Slosser and Milburn
\cite{SloMil94} considered electro-optic feedback of the photocurrent
from the idler mode of a non-degenerate parametric oscillator onto the
pump mode. Here the signal and idler mode formed the system and the
pump mode was the ancilla. The procedure they adopted for deriving a
master equation for the system was as follows. They expanded the
feedback master equation for the compound system to {\em first} order
in the feedback strength, adiabatically eliminated the pump mode, but
the final result presented for the system master equation
contained first and {\em second} order terms.
As in Ref.~\cite{WisMil94b}, this second-order master
equation was not of the Lindblad form. Furthermore, the
steady-state field averages were calculated using an unstated
{\em all}-order master
equation (which was of the Lindblad form). There are other problems
with this paper \cite{fn0}, but they are not relevant to the present work.

In this work we show how adiabatic elimination can be done rigorously
in compound quantum feedback systems such as those of
Refs.~\cite{WisMil94b,SloMil94}. As well as being of interest in the
field of quantum
feedback, the methods we use for adiabatic elimination are of more
general interest. While adiabatic elimination of an
ancilla mode
which is linearly coupled to the system is well understood, adiabatic
elimination with a nonlinear (e.g. proportional to the intensity) coupling
is not. In particular, the methods we use here put the results obtained by
Doherty and co-workers \cite{DohParTanWal98} on the motion of an atom
coupled to a damped optical cavity mode on a more rigorous
footing.

This paper is organized as follows. In Sec.~II we consider simple
direct-detection
feedback, and the four types of analogous feedback in compound
systems:
electro-optic feedback via a two-level atom, electro-optic feedback
via an optical mode, all-optical feedback via a two-level atom, and all-optical
feedback via a mode. We
show that in all four cases it is possible to eliminate the ancilla
under suitable conditions, giving a master equation for the
system alone. In Sec.~III we compare the stationary state of these
master equations with the solution of the full dynamics of the
compound systems. For this test we choose the free dynamics of the
system to be that of a
below-threshold parametric oscillator, the quantity being fed back to
be the intensity, and the quantity being
controlled by the feedback to be the  detuning.
We also compare the results of all five feedback
mechanisms
with that caused by an analogous ``reversible feedback''
generated by a $\chi^{(3)}$ nonlinearity.  In Sec.~IV we conclude
with a discussion of our results, and present a generalization
of the all-optical case to multiple optical modes.

\section{Adiabatic Elimination}

\subsection{Simple Feedback}

In order to discern how the dynamics of a system are affected by a
feedback loop that includes an ancilla, it is
useful to know the master equation for simple feedback.  By simple
feedback it is meant that the measurement results, based on
continuous observation of a source system, are immediately
used to alter the evolution of the source without the involvement of
any other quantum system.  To use an example from quantum optics, a
photodetector may register photon arrivals from a cavity
at discrete times and, at
these times, some specified change to the system may be made (see
Fig.~\ref{sfb2}).
 Types of changes include altering the optical path length or damping rate
 of  the cavity. In the remainder of this paper we will often use
 quantum optics terminology, but it should be remembered that the
 theory is not restricted to optical physics.

The most general form of the simple feedback master equation has
been derived by Wiseman \cite{Wis94a}.  Consider a system with
Hamiltonian $H$ and some dissipation at rate $\gamma$ and
with lowering operator $c$.
With $\hbar$ set equal to unity, the master equation is
\beq \label{me1}
\dot\rho(t) = -i[H,\rho] + \gamma{\cal D}[c]\rho,
\eeq
where the Lindblad \cite{lind} superoperator is
\beq
{\cal D}[c] = {\cal J}[c] -{\cal A}[c],
\eeq
where for arbitrary operators $A$ and $B$,
\beq
{\cal J}[A]B=A B A\dg\;;\;\;
{\cal A}[A]B =\half\{A\dg A,B\}.
\eeq
It is the dissipation which allows for continuous observation, the
result of which is a current $I(t)$. In this paper we are concerned
with what is known as direct detection where
\beq
I(t) = dN(t)/dt,
\eeq
where $dN(t)$ is the point process (the increment in the number of
photons counted) defined by
\bqa
[dN(t)]^{2} &=& dN(t) \nn \\
{\rm E}[dN(t)] &=& \gamma dt {\rm Tr}[c\dg c \rho_{\rm c}(t)]
\eqa
Here E denotes a classically probabilistic expectation value, while
the c subscript denotes that the state $\rho_{c}$ is conditioned on
the previous measurement results. We have assumed that the detection
is perfectly efficient; the generalization to inefficient detectors is
trivial \cite{Wis94a}.

Simple feedback arises from adding a Hamiltonian to the system
evolution of the form
\beq
H_{\rm fb}(t) = I(t) Z
\label{loopy}
\eeq
where $Z$ is an Hermitian system operator. Taking into account the singularity
of $I(t)$, and the fact that the feedback must act after the
measurement, it is possible to derive a master equation for the
system with feedback, averaging over all realizations of the
stochastic measurement record $I(t)$. The result is
\beq
\dot{\rho}= -i[H,\rho]+\gamma{\cal D}[e^{-iZ}c]\rho .
\label{eqnfbME}
\eeq
To compare this master equation with those obtained later it is useful
 to expand the exponentials to third order
\bqa
\dot{\rho} &\simeq &{\cal C}[H]\rho  +\gamma{\cal D}[c]\rho  \nn\\
&&+\;\gamma\left\{{\cal C}[Z]+\frac{1}{2}\left({\cal
C}[Z]\right)^{2}+\frac{1}{6}\left({\cal C}[Z]\right)^{3}\right\}
{\cal J}[c]\rho  ,
\label{3rdexpan}
\eqa
where ${\cal C}[A]B =-i[A,B ]$ for arbitrary operators $A$ and $B$.

The derivation outlined above for the feedback master equation treats
the photocurrent $I(t)$ as a classical stochastic process, which
causes the conditioned system state $\rho_{\rm c}$ to undergo
stochastic evolution (known as a quantum trajectory \cite{opsystem}).
There is an alternative derivation which treats the photocurrent
$I(t)$ as an operator. This derivation works in the Heisenberg
picture, where the system evolution is described by stochastic
operator differential equations known as quantum Langevin equations
\cite{inout}.
This method is useful for adiabatic elimination, so we will briefly
review its features.

Quantum Langevin equations (QLE) are constructed without
using the concept of measurement.
The dissipative evolution of \erf{me1} can be derived in a quantum
optical context from a linear coupling (in a rotating frame and with
the rotating wave approximation)
\beq
V = i\sqrt{\gamma}[v\dg(t) c - c\dg v(t)]
\eeq
between the system and a bath of harmonic oscillators.
Here $v(t)$ is the bath annihilation operator at the point at which
it interacts with the system. Just before this point, the bath is an
input vacuum, with field operator $v_{\rm in}(t)$ satisfying
\cite{inout}
\beq
[v_{\rm in}(t),v\dg_{\rm in}(t')] = \delta(t-t'),
\eeq
and all normally-ordered moments vanishing. Just after this point, the
bath is an output (non-vacuum) with field operator \cite{inout}
\beq
v_{\rm out}(t) = v_{\rm in}(t) + \sqrt{\gamma}c(t).
\eeq
The photocurrent operator $I(t)$ is simply the intensity of the output field
\beq
I(t) = v_{\rm out}\dg(t) v_{\rm out}(t).
\eeq

Adding together the evolution due to $H$, $V$, and $H_{\rm fb}$, and
again noting that the feedback must act after the interaction, one can
derive the following quantum Langevin equation for an arbitrary system
operator $s$ \cite{Wis94a}
\begin{eqnarray}
ds&=&[v_{{\rm
in}}^{\dag}+\sqrt{\gamma}c^{\dag}](e^{iZ}se^{-iZ}-s)[v_{{\rm
in}}+\sqrt{\gamma}c]dt\nonumber\\
&&+\gamma(c^{\dag }sc-\frac{1}{2}sc^{\dag }c-\frac{1}{2}c^{\dag
}cs)dt\nn\\
&&-\sqrt{\gamma}[dV_{{\rm in}}^{\dag
}c-c^{\dag }dV_{{\rm in}},s]+i[H,s]dt,
\label{eqnqlefb}
\end{eqnarray}
where $dV_{{\rm in}}= v_{{\rm in}}dt$.  All operators have time
argument $t$.
When the expectation value of
this equation is taken
an equation is obtained that can be converted to the master equation
(\ref{eqnfbME}) for simple feedback. If $Z$ is set to zero then the
Langevin equation describes damping alone.

\subsection{Electro-optic Feedback via an Atom}
 \label{BB}

The simplest possible ancilla system is a two-level atom (TLA).
In this section we consider incoherent (electro-optic)
feedback via this ancilla. The output from the  system is monitored by
direct detection, the results of which are used to affect the
evolution of the two level atom which is coupled to the system, as
shown in Fig.~\ref{eoat2}.
The system and ancilla are assumed to have
approximately the same resonant frequency.
If the atom is to be adiabatically eliminated, it must be heavily
damped, in which case it will mostly be in the ground state. Then
the most natural
form of feedback involves flipping the state of the TLA whenever the
photodetector monitoring the system makes a detection.  This can be
achieved with a feedback Hamiltonian of the form
\beq
H_{{\rm fb}}=\frac{\pi}{2}\sigma_{x}I(t).
\eeq
Here $\sigma_{x}$ is the usual Pauli spin matrix for describing an
atomic state \cite{pauli}.
It could be realized experimentally by very briefly driving the atom
with a pulse of on-resonance radiation (a `$\pi$' pulse) which will
flip it from the ground to the excited state.

With this form of feedback, the obvious
coupling of the atom to the system to consider is one proportional
to the excited state population operator $\sigma\dg \sigma$.
Here $\sigma = (\sigma_{x}-i\sigma_{y})/2$ is the atomic lowering
operator.
Specifically,
\begin{equation} \label{Hcoup3}
H_{{\rm coupling}}=\sigma^{\dag}\sigma K,
\label{sigma+1}
\end{equation}
where $K$ is an arbitrary Hermitian system operator.
When feedback onto the atom in
the ground state occurs the upper state population jumps to a value of 1
and then
decays away, due to coupling to the continuum of electromagnetic
field modes.  In other words, $\sigma\dg \sigma$ will tend to follow
the photocurrent. Thus there is a strong
similarity to simple feedback, if $K$ is chosen
to be some scalar
multiple of $Z$.

It is not hard to generalize \erf{eqnfbME} to include the
TLA ancilla
\bqa
\dot{W}&=&-i[H_{{\rm system}}
+\sigma^{\dag}\sigma K,W]\nn\\
&&+{\cal D}[\exp (-i\frac{\pi}{2}\sigma_{x})c]W+\Gamma {\cal
D}[\sigma ]W,
\eqa
where $\Gamma$ is the damping rate of the atom and $W$ is the
density matrix for the compound system.  The damping rate $\gamma$
of the system
has been set equal to unity without loss of generality.
  Of course, the operators
are now acting in the joint
Hilbert space of the two systems so that $c\equiv c\otimes
1_{{\rm atom}}$ and $\sigma\equiv 1_{{\rm
system}}\otimes\sigma$, etc.

The above master equation gives the evolution of the density operator
for the compound system.  At any time a partial trace of
this operator over the atom could be performed to obtain the reduced
density matrix for the  system alone.  However, in general,
this cannot be done to the master equation itself in order to obtain
a master equation for $\rho_{{\rm system}}(t)$.  The obvious exception to this
is the case where $K=0$ and the system is
unaffected by the atom.

It is logical that a master equation for the system  cannot be
derived if
the atom observables
fluctuate, in response to the feedback, on the same time scale as the
system observables.  The effect of feedback would then depend on the
constantly fluctuating state of the atom which, in turn, depends on
previous feedback.  Removing the atom operators from the master
equation without removing information concerning the system is
impossible due to the coupling that exists between them.  Of course,
a non-Markovian
expression could be written down for the atom in terms of the system,
but this would not lead to a Lindblad master equation without some further
approximation.

If the atom reacts very quickly to the feedback and returns
to its initial state before more feedback arrives (the next
photodetection) then this well defined behavior can be built into a
master equation for the system alone. In essence, the atom's state is
approximated by its equilibrium value with respect to the
instantaneous state of the system and operators are replaced by their
steady state expressions.  This procedure is known as
adiabatic elimination of the
atom.

To proceed with the adiabatic elimination it is noted that the total
density matrix can be expanded as
\bqa
W=&\rho_{0}\otimes \ket{\downarrow}\bra{\downarrow}
+\rho_{1}\otimes \ket{\downarrow}\bra{\uparrow}\nn\\
&+\rho_{1}^{\dag}\otimes\ket{\uparrow}\bra{\downarrow}+
\rho_{2}\otimes\ket{\uparrow}\bra{\uparrow},
\label{rhoexpan}
\eqa
where the $\rho$s exist in the system subspace.  All possible
states of the atom have been included ($\ket{\uparrow}$ and
$\bra{\downarrow}$ correspond to the excited and ground state
respectively).  This approach is particularly appropriate because of
the small basis involved.  If the above expression for $W$ is
substituted into the master equation then the atom operators can act
on the states of the atom.  If the coefficients of the various
orthogonal states are equated the following equations for the $\rho
$s are obtained (the subscript `s' indicates the system):
\begin{eqnarray}
\dot{\rho_{0}}&=&{\cal C}[H_{\rm s}]\rho_{0}+{\cal J}[c]\rho_{2}-{\cal
A}[c]\rho_{0}+\Gamma\rho_{2},\\
\dot{\rho_{1}}&=&{\cal C}[H_{\rm s}]\rho_{1}+i\rho_{1} K+{\cal J}[c]\rho^{\dag}_{1}-{\cal
A}[c]\rho_{1}-\frac{\Gamma}{2}\rho_{1},\\
\dot{\rho_{2}}&=&{\cal C}[H_{\rm s}]\rho_{2}+{\cal
J}[c]\rho_{0}-{\cal A}[c]\rho_{2}-\Gamma\rho_{2}.
\label{eqnrho2dot}
\end{eqnarray}
By tracing
\erf{rhoexpan} over the atom the reduced density operator
for the system is
\begin{equation}
\rho_{{\rm s}}=\rho_{0}+\rho_{2}
\end{equation}
and its evolution equation is found to be
\begin{equation}
\dot{\rho}_{{\rm s}}=-i[H_{{\rm s}},\rho_{{\rm
s}}]+{\cal D}[c]\rho_{{\rm s}}-i[K,\rho_{2}].
\label{eqndotrhocav}
\end{equation}

Without some approximation this is as far as the elimination of the atom
can be taken.  It is not a master equation due to the dependence upon
$\rho_{2}$.  As discussed previously, the limit in which the atom
returns very quickly to the ground state after feedback needs to be
considered.
Because the probability
for photodetection in any infinitesimal time period scales as the
size of the period, the atom is in the ground state for
almost all time.  The approximation that $\rho_{{\rm
s}}\approx\rho_{0}$ is therefore made.  To obtain a master
equation, an expression for $\rho_{2}$ in terms of $\rho_{0}$ is
needed.  From \erf{eqnrho2dot} it can be seen that if
$\Gamma$ is large compared to the other co-efficients of
$\rho_{2}$ (except possibly $K$) then fluctuations in this operator
will be quickly damped out and $\dot{\rho_{2}}$ can then be set to
zero.  The effect of $K$ is to cause rotation of
$\rho_{2}$ but not to affect its size.
The physical picture already described is consistent with $\Gamma$ being
large.
Assuming  $K\sim\Gamma\gg 1$ (where $K \sim\Gamma$ means that the operator
$K$ scales like $\Gamma$),
we find the steady state of $\rho_{2}$ to be
\begin{equation}
\rho_{2}=(\Gamma -{\cal C}[K])^{-1}{\cal J}[c]\rho_{0}.
\end{equation}
When this is
substituted into \erf{eqndotrhocav} the master equation for
the system alone is obtained.  With $Z=K/\Gamma$ it is
\bqa
\dot{\rho}_{{\rm s}}= \{ {\cal C}[H_{\rm s}]
+{\cal D}[c]+
{\cal C}[Z](1 -{\cal C}[Z])^{-1}{\cal
J}[c]\}\rho_{{\rm s}}.
\label{hello}
\eqa
It is not immediately clear that
this master equation is of the Lindblad form \cite{lind}.
However in appendix~\ref{APPlindtla} it
is shown that it can be written as
\begin{equation}
\dot{\rho}_{{\rm s}}=-i[H_{{\rm s}},\rho_{{\rm s}}]+\int_{0}^{\infty}dq
e^{-q}{\cal
D}[e^{-iqZ}c]\rho_{{\rm s}}.
\label{berry}
\end{equation}

Some feeling for the nature of the master equation can be obtained by
an expansion to third order in $Z$ (a small feedback
approximation).  This gives (subscripts dropped)
\bqa
\dot{\rho}&\simeq&
{\cal C}[H]\rho+{\cal
D}[c]\rho\nn\\
&&+\;\left\{{\cal C}[Z]+({\cal C}[Z])^{2}+({\cal C}[Z])^{3}\right\}{\cal J}[c]
\rho .
\label{eqntla}
\eqa
These terms can be compared to the third order expansion of
\erf{3rdexpan}, with $\gamma =1$.  The difference in
second and higher order terms
means that for large feedback the two systems will be significantly
different.

\subsection{Electro-optic Feedback via a mode}
\label{elecmode}
The more challenging task of adiabatically eliminating an ancilla
that has an infinite number of basis states is now considered.
Optically, this could correspond to a single-mode cavity. The
method of expanding the compound density matrix in terms of the
lower number states of the ancilla is not appropriate due to
the type of feedback that is utilized.  Instead we use
Quantum Langevin equations, which place no such restriction on the
excitation of the ancilla.

The output
field from the system is once again continuously monitored using
direct detection (see Fig.~\ref{eocav2}).
We take the feedback to
be linear driving of the ancilla cavity.
This causes a jump in amplitude of the ancilla cavity
 when there is a photodetection. It is described by the
  feedback Hamiltonian
\begin{equation}
H_{{\rm fb}}=\frac{\epsilon}{2}(-ib+ib^{\dag})I(t),
\end{equation}
where $b$ is the annihilation operator for the cavity, $\epsilon$
represents the amplitude of the coherent driving field and $I(t)$ is
the operator for the photocurrent output from the system.  Its effect
can be determined from the Heisenberg equation of motion for $b$,
\beq
\dot{b}_{{\rm fb}}=-i[b,H_{{\rm fb}}]=\frac{\epsilon}{2} I(t).
\label{eqnbfb}
\eeq
Since $I(t)$ consists of $\delta$ functions, it is
clear that the cavity field amplitude changes by an amount
$\epsilon/2$ whenever a photodetection occurs.  Note that here
the implicit equation of motion for $b$ is sufficient to determine
its evolution because the stochastic term is not dependent upon $b$
\cite{qnoiseg}.
To provide a feedback circuit that is classically equivalent to
simple feedback in the limit of large damping of the cavity, the
following choice of coupling is made:
\begin{equation}
V=\frac{K}{2}(b+b^{\dag}).
\end{equation}
The equivalence can be seen if linear damping is included in \erf{eqnbfb}.
The slaved value of $b$ (in the limit of large damping $\dot{b}_{{\rm
fb}}$ is set equal to zero) is then substituted into the
coupling, which leaves it in the same form as a simple feedback
Hamiltonian, given an appropriate choice of $K$.

The total master equation is
\bqa
\dot{W}&=&-i[\frac{K}{2}(b+b^{\dag})+H_{{\rm s}},W]\nn\\
&&+{\cal D}[e^{\epsilon (-b+b^{\dag})/2}c]W+\Gamma{\cal D}[b]W,
\label{eqnMEtwocav}
\eqa
where once again $W$ is the density matrix describing the compound
system and the damping of the system has been set equal to unity.
The damping rate of the ancilla cavity is given by $\Gamma$.  The quantum
Langevin equation that corresponds to this master equation can be
found by extending \erf{eqnqlefb}.  The result for an
arbitrary operator $r$ from either sub-system is
\bqa
&dr=&v_{{\rm out}}^{\dag}\left[e^{
\epsilon (b-b^{\dag})/2}re^{
-\epsilon (b-b^{\dag})/2}-r\right]v_{{\rm out}}dt\nn\\
&&+{\cal D}[c^{\dag}]rdt-[dV_{{\rm in}}^{\dag}c-c^{\dag }dV_{{\rm
in}},r]\nn\\
&&+\Gamma {\cal D}[b^{\dag}]rdt-\sqrt{\Gamma}
[dU_{{\rm in}}^{\dag}b-b^{\dag }dU_{{\rm
in}},r]\nn\\
&&+i[\frac{K}{2}(b+b^{\dag})+H_{{\rm s}},r]dt,
\label{eqndstwocav}
\eqa
where $dU_{{\rm in}}=u_{{\rm in}}dt$.  The vacuum
field input for the driven cavity, $u_{{\rm in}}$, has the same properties as
$v_{{\rm in}}$.

To adiabatically eliminate the cavity, in the limit of
heavy damping, a QLE will first be determined for a system operator,
$s$.  \erf{eqndstwocav} is greatly simplified, as $s$
commutes with all driven cavity operators, to give
\begin{eqnarray}
ds&=&{\cal D}[c^{\dag}]sdt-[dV_{{\rm in}}^{\dag}c-c^{\dag }dV_{{\rm
in}},s]\nonumber\\
&&+i[\frac{K}{2}(b+b^{\dag})+H_{{\rm s}},s]dt.
\end{eqnarray}
From this it is evident that an expression for $b$ is
required if a master equation for the system alone is to be derived.
The QLE for $b$ is
\begin{equation}
\dot{b}=-i\frac{K}{2}-\frac{\Gamma}{2}b-\sqrt{\Gamma}u_{{\rm in}}
+\frac{\epsilon}{2} v_{{\rm out}}^{\dag}v_{{\rm out}}.
\end{equation}
For large $\Gamma$ the fluctuations in $b$ due to system
operators will be quickly damped out.  However, the stochastic terms
have an infinite bandwidth, so that it is not strictly possible to
slave an operator that only responds to a finite bandwidth,
$\Gamma$, to these fluctuations.
Although this problem can be side-stepped \cite{WisMil94b} it will
prove advantageous to use the following equilibrium value of $b$
\bqa
b&=&-\frac{iK}{\Gamma}-\int_{0}^{\infty}d\tau e^{-\Gamma\tau /2}
\left[\sqrt{\Gamma}u_{{\rm
in}}(t-\tau) \phantom{\frac12}\right. \nn\\
&&\phantom{-\frac{iK}{\Gamma}+\int_{0}^{\infty}d\tau }
\left. -\;\frac{\epsilon}{2} v_{{\rm out}}^{\dag}v_{{\rm
out}}(t-\tau)\right] .
\eqa
The integral serves to determine the present contribution to $b$ from
the stochastic terms at time $t-\tau$. This contribution falls off at
rate $\Gamma/2$, the amplitude
decay rate for the ancilla cavity. The term that is not under the
integral comes from $K$ which is not stochastic and is therefore
slowly varying compared to the highly damped
cavity operators.  Thus, $b$ can follow its evolution to a very good
approximation.

To simplify matters the Langevin equation for $s$ will now be
rearranged before substitution so that $u_{{\rm in}}$ will annihilate
the vacuum when the expectation value is taken.  This gives
\begin{eqnarray}
ds&=&{\cal D}[c^{\dag}]sdt-[dV_{{\rm in}}^{\dag}c-c^{\dag }dV_{{\rm in}},s]
\nonumber\\
&&+\,\frac{i}{2}\left(b^{\dag}[K,s]+[K,s]b\right)dt+i[H_{{\rm
s}},s]dt.
\end{eqnarray}
This is valid as $b$ and $b^{\dag}$ commute with system operators.
We cannot move
the stochastic part $v_{\rm in}(t)$ of
$v_{{\rm out}}(t-\tau )$ through the system commutator term to
annihilate on the vacuum.  However, it is
possible to move the photocurrent itself at time $t-\tau$ as it
commutes \cite{inout}.
If the integrals that will annihilate on the vacuum when the trace
over the bath is taken are ignored, then we are left with
\bqa
\dot{s}&=&\frac{i\epsilon}{2} \left[K,s\right]\int_{0}^{\infty}d\tau
e^{-\Gamma\tau /2}I(t-\tau) -\frac{1}{2\Gamma}\left[
K,[K,s]\right]\nn\\
&&+{\cal D}[c^{\dag}]s
-[v_{{\rm in}}^{\dag}c-c^{\dag }v_{{\rm in}},s]
+i[H_{{\rm s}},s].
\label{implicit}
\eqa
If the limit $\Gamma\rightarrow\infty$ is taken the
integral reduces to $2I(t)/\Gamma$.  The resultant equation for $\dot{s}$
is an implicit equation as it was derived by idealizing the
properties of the cavity and environment \cite{qnoiseg}.  An explicit
equation is
now required.

The term that needs to be treated in \erf{implicit} can be
written as
\begin{equation}
\dot{s}_{{\rm implicit}}=-\frac{\epsilon I{\cal C}[K]s}{\Gamma}.
\label{eqnsdotterm1}
\end{equation}
This gives an explicit increment of the form \cite{Wis94a,handbook}
\begin{equation}
ds_{{\rm explicit}}=dN[\exp (-\epsilon{\cal C}[K]/\Gamma)-1]s,
\end{equation}
where $dN=Idt=dN^{2}=v_{{\rm out}}^{\dag}v_{{\rm out}}dt$.
Remembering that the photocurrent is actually evaluated at a slightly
earlier time than the system operators allows $v_{{\rm out}}$ to be
moved to the right of the expression. If we put $Z=\epsilon K/\Gamma$, in
order
that our equations can be compared to simple feedback, then
the total Langevin equation is
\bqa
ds&=&[v_{{\rm in}}^{\dag}+c^{\dag}]
(e^{iZ}se^{-iZ}-s)[v_{{\rm in}}+c]dt\nn\\
&&-\frac{\Gamma}{2\epsilon^{2}}\left[
Z,[Z,s]\right]dt
+{\cal D}[c^{\dag}]sdt\nn\\
&&-[dV_{{\rm in}}^{\dag}c-c^{\dag }dV_{{\rm in}},s]+i[H_{{\rm s}},s]dt.
\eqa
When the expectation value is taken the stochastic part annihilates
on the
vacuum and the following master equation is obtained
\begin{equation}
\dot{\rho}=-i[H_{{\rm s}},\rho]+
{\cal D}[e^{-iZ}c]\rho +\frac{\Gamma}{\epsilon^{2}}
{\cal D}[Z]\rho.
\label{eqnMEadtwocav}
\end{equation}
The only difference from simple feedback is the third term.
This is a term of second order in the feedback operator $Z$,
and  represents a type of noise that will tend to smooth
over the interesting behavior of the system.  Clearly it can be made
arbitrarily small if $\epsilon$ is made large enough.
A more detailed
discussion of this term is given in Sec.~\ref{blahbar}

\subsection{All-optical Feedback via an atom}

We turn now to coherent, or all-optical feedback.
Once again we begin with the simplest possible ancilla, a
two-level atom.
All-optical feedback via an atom involves the reflection of the output
field from the system onto the atom, where the atom is reversibly coupled
to the
system.  Here, the resonant frequencies of the two systems are taken
to be equal.  It is different from electro-optic feedback as there is no
measurement ste; the light is just reflected around a loop with the
use of mirrors (see Fig.~\ref{alloptat2}).
The theoretical description of such
systems was developed largely by Carmichael \cite{copentheory2} and
Gardiner \cite{copentheory} and has been termed {\em Cascaded Open
Systems theory}.  If linear bath-system couplings are assumed then
the compound master equation is
\bqa
\dot{W}=&-i[H_{\rm s}+V,W]+{\cal D}[c]W+\Gamma{\cal
D}[\sigma ]W\nn\\
&+\sqrt{\Gamma}\left([cW,\sigma^{\dag
}]+[\sigma,Wc^{\dag }]\right).
\label{eqnMEallopt}
\eqa
The system damping has been set equal to unity as usual
 and $\Gamma$ is the damping rate of the
atom.

In order to investigate the degree to which all-optical
feedback can replicate electro-optical simple feedback, a coupling is
chosen that is linear in the excited state population
of the atom.  We expect this operator to follow the output
photocurrent from the system.  That is, we assume a coupling
\begin{equation}
V=K\sigma^{\dag}\sigma
\label{qed}
\end{equation}
identical to that in Sec.~\ref{BB}.
 Making the
expansion of \erf{rhoexpan} gives the following for the
$\rho$s
\bqa
\dot{\rho_{0}}&=&{\cal C}[H_{\rm s}]\rho_{0}+{\cal D}[c]\rho_{0}
+\Gamma\rho_{2}+\sqrt{\Gamma}(c\rho_{1}+\rho_{1}c^{\dag}),\\
\dot{\rho_{1}}&=&{\cal C}[H_{\rm s}]\rho_{1}+{\cal D}[c]\rho_{1}+\sqrt{\Gamma}
(\rho_{2}-\rho_{0})c^{\dag}\nn \\
&& +\,i\rho_{1}K-\frac{\Gamma}{2}\rho_{1},\\
\dot{\rho_{2}}&=&{\cal C}[H_{\rm s}]\rho_{2}+{\cal D}[c]\rho_{2}
-\sqrt{\Gamma}(c\rho_{1}+\rho_{1}c^{\dag})\nn\\
&&-\,i[K,\rho_{2}]-\Gamma\rho_{2}.
\eqa
  The
above equations lead to an equation of motion for the system density
operator of
\beq
\dot{\rho}={\cal C}[H_{\rm s}]\rho +{\cal D}[c]\rho -i[K,\rho_{2}],
\label{atalopt}
\eeq
which is the same as \erf{eqndotrhocav}.  To find an expression
for $\rho_{2}$ the normal procedure of taking
$\Gamma$ large compared to ${\cal C}[H_{\rm s}]$ is performed.
Thus, $\rho_{1}$ can be slaved to system operators, $\rho_{0}$ and
$\rho_{2}$.  Now as we only require a master equation which gives the
leading order effect in $\Gamma^{-1}$ of
the ancilla on the system, $\rho_{2}$
can be set equal to zero in the $\rho_{1}$ equation, which is the
approximation $\rho_{0}\approx\rho$.  This is valid
as $\rho_{2}\sim\rho_{0}/\Gamma$.
By substituting
the slaved expression for $\rho_{1}$ into that for $\rho_{2}$ we find
after simplification
\beq
\rho_{2}=\frac{4}{\Gamma}{\cal
J}\left[\left(1+\frac{2iK}{\Gamma}\right)^{-1}c\right]\rho_{0}.
\eeq
This can now be substituted into \erf{atalopt} to obtain a master
equation.  Writing $Z=4K/\Gamma$, we have
\begin{equation}
\dot{\rho}={\cal C}[H_{\rm s}]\rho +{\cal D}[c]\rho+{\cal C}[Z]{\cal
J}\left[\left(1+\frac{Zi}{2}\right)^{-1}c\right]\rho ,
\label{eqnadallopt}
\end{equation}
which is the same as the simple feedback \erf{3rdexpan} to second order.
 The third order
term is
\beq
\frac{1}{4}{\cal C}[Z]({\cal J}[Z]-2{\cal A}[Z]){\cal J}[c]\rho .
\eeq
Again it is not obvious that \erf{eqnadallopt} is in the Lindblad
form, but it is shown in
Appendix~\ref{aloptlind} that it can be written
as
\begin{equation}
\dot{\rho}=-i[H_{{\rm
s}},\rho]+{\cal D}\left[\exp
\left(-2i\arctan{\frac{Z}{2}}\right)c\right]\rho.
\label{g'daymate}
\end{equation}

\subsection{All-optical Feedback via a mode}
 \label{ee}
The final compound system that will be considered involves the output
field from a system being reflected onto an optical cavity
 that is coupled
back to the system (see Fig.~\ref{alloptcav2}).
A Faraday Isolator (comprised of a Faraday Rotator and a Polarization
dependent
Beam Splitter) prevents reflected light from the
cavity returning to the system.
  The only difference in the total master equation from
the previous section is the replacement of the atom lowering
operator $\sigma$ with the annihilation operator $b$.  Thus a coupling of the
form $V=Kb^{\dag}b$ is considered.

The derivation of a master equation for the system alone
follows similar lines to that of Sec.~\ref{elecmode}.  The QLE for
an arbitrary operator is \cite{copentheory}
\begin{eqnarray}
dr&=& +i[H_{{\rm s}}+V,r]dt+{\cal D}[c^{\dag}]rdt-[dV_{{\rm
in}}^{\dag}c-c^{\dag }dV_{{\rm in}},r]\nn\\
&&+\Gamma {\cal D}[b^{\dag}]rdt-\
\sqrt{\Gamma}[dV_{{\rm in}}^{\dag}b-b^{\dag }dV_{{\rm in}},r]\nonumber \\
&&+\sqrt{\Gamma}(b^{\dag }rc+c^{\dag }rb-rb^{\dag }c-c^{\dag
}br)dt.
\label{eqndsallopt}
\end{eqnarray}
For a system operator this becomes
\bqa
ds&=&{\cal D}[c^{\dag}]sdt-[dV_{{\rm
in}}^{\dag}c-c^{\dag}dV_{{\rm in}},s]\nn\\
&&+\, i[H_{{\rm s}}+Kb^{\dag}b,s]dt.
\label{eqndssourceallopt}
\eqa
The next step is to find an equation for $b$.  The QLE that governs
it is
\begin{equation}
db=-(\frac{\Gamma}{2}b+\sqrt{\Gamma}v_{{\rm
in}}+\sqrt{\Gamma}c+iKb)dt.
\label{eqndotballopt}
\end{equation}
This justifies our initial presumption that the cavity photon number would
follow the photocurrent.  For $\Gamma$ large it is
possible to slave $b$ to the system operators and to form an
integral expression for the contribution from the stochastic term, as
in Sec.~\ref{elecmode}. The result is
\bqa
b&=&-\frac{2}{\sqrt{\Gamma}}\left(1+\frac{2iK}{\Gamma}\right)^{-1}c\nn\\
&&+\sqrt{\Gamma}\int_{0}^{\infty}d\tau
e^{-\Gamma\tau(1+2iK/\Gamma )/2}v_{{\rm in}}(t-\tau).
\label{retro}
\eqa
The same trick of rearranging the QLE for the system operator is again used
so that, in this case, all of the integral terms annihilate.  We put
\begin{equation}
i[Kb^{\dag}b,s]dt=ib^{\dag}[K,s]bdt.
\label{trick}
\end{equation}
Substituting into this the expression for $b$ and $b^{\dag}$ gives
four terms, only one of which is non-zero when the trace over the bath
is taken.  This term is
\begin{equation}
\frac{4ic^{\dag}}{\Gamma}\left(1-\frac{2iK}{\Gamma}\right)^{-1}
\left[K,s\right]\left(1+\frac{2iK}{\Gamma}\right)^{-1}c.
\label{eqndscouplingterm}
\end{equation}
In effect, an implicit equation has been derived that has no
contribution from stochastic operators, resulting in there being no
need for an implicit/explicit distinction.  It is now possible to
turn the equation for $ds$ into a master equation for the system.  When
this is
done we arrive at the same result as \erf{eqnadallopt}.
The conclusion is that to first order in $\Gamma^{-1}$, the cavity
has the same effect on the system that the atom does, when included in
an all-optical feedback loop.

In hindsight, this is what we should have expected,
as in the limit of large damping only the lowest number states of
the cavity will be occupied with significant probability. One
could therefore have expanded the total density matrix analogously to
the TLA system to obtain the same equations immediately.

The reason why
electro-optic feedback onto an atom and a cavity were not equivalent is
due to the more singular nature of the driving of the ancilla.  When a
detection on the system is made the field amplitude of the
cavity jumps, leading to occupation of higher photon number states.
These states are,
therefore, essential to the description of the compound system.
Electro-optic feedback
onto the atom cannot replicate this behavior.

\section{COMPARISON WITH EXACT RESULTS}

We have shown that in principle it is possible to consider a variety
of different sorts of feedback
in compound quantum systems, and to adiabatically eliminate the
ancillary system to arrive at master equations for the system of
interest alone. These master equations should be exact in the limit
that the ancilla is damped infinitely faster than the system. In
practice, this will never be the case, so it is an interesting
question to find out under what conditions the equations are valid.
This can be done by simulating the full master equation for the compound
system and comparing to the results of the master equation for the
system alone.

To make such a comparison requires specifying the
feedback operator, $Z$,  and the system Hamiltonian, $H_{{\rm s}}$.
Once this is done, a  comparison can be made by looking at the stationary
solutions of the respective master equations. While this could be
criticised as not being a complete test, it has the advantages of
definiteness and ease of calculation (in some cases at least).
Furthermore,  we choose a system (a damped optical mode) and
Hamiltonians $H_{\rm s}$ and $Z$ such that the stationary solutions
have enough structure for the comparison to be interesting.
The comparison is both quantitative and qualitative, with the use of the
Bures distance
\cite{Bures} as a measure of the difference between
the state
matrices and the Wigner function \cite{wallsmilburn} to illustrate them.

In the hope of getting some interesting states
we take the  system to be a damped single mode optical cavity. That
is, we choose $c=a$, an annihilation operator satisfying $[a,a\dg]=1$.
We choose a system Hamiltonian (in a rotating frame) of
\beq
H=-\frac{i\lambda}{4}\left[a^{2}-(a^{\dag})^{2}\right],
\eeq
This describes a degenerate parametric amplifier  (``two photon"
driving), which can be realized by driving
an intracavity crystal with a $\chi^{(2)}$ non-linearity
with light at twice the resonant frequency.
For $\lambda$ positive,
this results in squeezing of the $X_{2}$ quadrature  of the
field inside the cavity, and stretching of the $X_{1}$ quadrature.
The two
 quadratures are defined in this paper as
\bqa
X_{1}&=&a+a^{\dag}\\
X_{2}&=&-i(a-a^{\dag}).
\eqa
Without feedback, the master equation with two-photon driving and
damping will have a stationary solution only for $\lambda < 1$.
That is, $\lambda$ is the threshold parameter.

The feedback operator is chosen to be
\beq
Z=\chi a^{\dag}a.
\eeq
We can get a feel for the effect of this type of feedback by using
$Z$ in the simple feedback Hamiltonian given in \erf{loopy}.  That
is,
\beq
H_{{\rm fb}}(t)=\chi I(t) a^{\dag}a.
\label{newday}
\eeq
This represents a detuning of the system cavity proportional to the
photocurrent. It will cause the master equation to have
a stationary solution
 regardless of $\lambda$, as will be shown.  As the mean photocurrent is
 equal to the expectation value
of the photon number operator for the system, this Hamiltonian is
akin to a $\chi^{(3)}$ Kerr non-linearity \cite{wallsmilburn}.
  In Sec.~\ref{nonlin} a
comparison of feedback to such a nonlinearity is made.

\subsection{Simple Feedback}

The master equation for simple feedback is now
\beq
\dot{\rho}=-\frac{\lambda}{4}\left[a^{2}-(a^{\dag})^{2},\rho\right]+{\cal
D}[e^{-i\chi a^{\dag}a}a]\rho .
\eeq
To simplify the numerical analysis we
 choose a single feedback strength for which simulations will be run.
To aid this decision the effect of feedback is analyzed.  Consider
the following quantity:
\bqa
{\cal J}\left[e^{-i\chi a^{\dag}a}\right]\rho .
\eqa
If this is evaluated in the number basis then we get
\beq
\bra{n}{\cal
J}\left[(e^{-i\chi})^{a^{\dag}a}\right]\rho\ket{m}=(e^{-i\chi})^{n-m}\rho_{
nm}.
\eeq
Now this particular system has the property that $\rho_{nm}=0$ for
$|n-m|$
odd as the
two photon driving is the only source of coherences.  These coherences
exist between elements with $|n-m|$ even.  Hence, if $\chi =q\pi$, with
$q$ an integer, then
the feedback has no effect.  Investigation into the states produced with
a value of feedback close to this revealed that they are extremely
sensitive to
any parameter variation.  This implies that it is not a suitable
regime for the testing of adiabatic elimination.  The most obvious
alternative is to choose the maximum feedback regime.  It is clear
that this is achieved with $\chi =(q+1/2)\pi$.  The states produced are
much less sensitive and also have the advantage that, for simple
feedback, there is no
threshold to the driving strength above which the photon number
becomes infinite. For the remainder of
the paper we choose $\chi=\pi/2$.

The two-photon driving strength $\lambda$ was chosen to be as large
as possible, given the constraints on the maximum basis size that
could be simulated. This amplified the interesting effects of feedback.
Not surprisingly, the simulations of the compound
systems are the most computationally intensive and provide the upper
basis size.  It was found that the limit for the system cavity basis
size required that photon numbers
above 35 had to be truncated.  For an accurate simulation
\cite{endnote}
 this gives a maximum driving strength of
about $\lambda =2.2$. Where possible,
the compound systems were examined in the same regime as
simple feedback, but for some the driving threshold of $\lambda=1$
remains in force, so $\lambda =0.97$ was then chosen.

The numerical simulations were greatly aided by the use of the Quantum
Optics toolbox for Matlab \cite{tool}.  As noted above,
we gauged  whether the adiabatic
elimination is valid by investigating the steady states of the systems.
 The simple feedback system involved a small enough
Liouvillian that matrix inversion methods can be used.  The Wigner
function of the steady state density matrix for simple feedback, with
$\lambda =2.2$ and $\chi =\pi /2$, is shown in
Fig.~\ref{sfbsubplot}.  A plot with $\lambda =0.97$ is also included.

\subsection{Electro-optic Feedback via an Atom}

Electro-optic feedback via an atom can be compared
to the simple feedback
just considered if we insert in \erf{Hcoup3}
 $K= \Gamma Z = g a^{\dag}a$, where $g=\Gamma \pi /2$.  To test the
adiabatic
elimination simulations were run for various values of
$\Gamma$.  It is only for large $\Gamma$ that correspondence between the
full dynamics and the adiabatically eliminated master equation is
expected.
 A physical realization of this coupling is a far
detuned atom in the standing wave of a single mode
 cavity \cite{wallsmilburn}.  This also introduces a term
into the system Hamiltonian of the form $\delta\sigma^{\dag}\sigma$,
where $\delta$ is the difference in resonant frequency of the
atom and  system.  It
is of interest to determine whether the same results are obtained
if the adiabatic elimination is done at the same time, rather than
after, the large-detuning approximation is made.  This is addressed in
appendix~\ref{JC}, and the answer is affirmative.

The full
master equation is of the form
\bqa
\dot{W}&=&-i\left[-\frac{i\lambda}{4}\left\{a^{2}-(a^{\dag})^{2}\right\}
+g\sigma^{\dag}\sigma a^{\dag}a
+\delta\sigma^{\dag}\sigma,W\right]\nn\\
&&+{\cal D}[\exp (-i\frac{\pi}{2}\sigma_{x})a]W+\Gamma {\cal
D}[\sigma ]W.
\eqa
The reduced density matrix for the system at steady state needs to be
found.  Once again, the Liouvillian is small enough that we can set
$\dot{W}=0$ and solve the equation ${\cal L}W=0$ for the non-trivial
solution.  Simulations were run for values of $\Gamma$ from 1 to 100, with
$g$ altered accordingly.  Note that the detuning actually has no effect on
the system dynamics.
The reduced density matrices produced are
compared with those found from \erf{hello} with the aid of the Bures
distance, which gives a measure of how distinguishable two mixed
states ($\rho_{1}$ and $\rho_{2}$)
are.
The Bures distance is defined as \cite{bures1}
\beq
d_{{\rm Bures}}(\rho_{1},\rho_{2})=\sqrt{2}\left(1-{\rm Tr}
\left[\sqrt{\sqrt{\rho_{1}}\rho_{2}\sqrt{\rho_{1}}}\right]\right).
\eeq
All pairs of density matrices of the same size have a Bures measure
that is mapped onto the real numbers between zero and $\sqrt{2}$.
Fig.~\ref{eoatsubplot} shows how the state produced by the compound
master equation
approaches that produced by the adiabatically eliminated master equation.
As $\Gamma$ is increased the Bures distance decreases and the Wigner
functions become more similar to the adiabatic state.
This shows that the adiabatic elimination is
valid in this system for surprisingly small values of $\Gamma$.

A comparison of the stationary Wigner functions produced here with those of
simple
feedback reveals that there exists vast differences between these
feedback schemes.  This is not surprising as it is only to first
order in $Z$ that the equations are the same, and
the parameters we have chosen correspond to $Z$ quite large.
The most obvious visual differences include the presence of a shearing effect
and the
loss of reflective symmetry
in the $X_{2}$ quadrature.

\subsection{Electro-optic Feedback via a Mode}
\label{blahbar}
In Sec.~\ref{elecmode} electro-optic feedback via a mode was
considered.  In
the limit of the ancilla mode being damped on a time scale small compared to
those of the
system, \erf{eqnMEadtwocav} was derived.
The feedback operator was set as $Z=\epsilon K/\Gamma$ so that we
could make a comparison to simple feedback. It follows that
 the system coupling operator, $K$, is of the same form as the
previous section: $K=ga^{\dag}a$.  The coupling
$V=ga^{\dag}a(b+b^{\dag})/2$ could be physically achieved
via a four wave mixing process in a $\chi^{(3)}$ material
\cite{wallsmilburn}. The
fourth
field would have to have the same frequency as the ancilla cavity for
conservation of energy.

Now that $Z$ has been specified,
 the third term in \erf{eqnMEadtwocav}
 can be discussed more explicitly.  This can be done by considering
the evolution of the phase operator, which has an approximate commutation
relation with the number operator of $[\Phi ,n]=-i$ \cite{dirac}.  It can
then be
shown that this term causes phase diffusion at a constant rate, implying that the features
of the state which are dependent upon a distinct phase are lost.
With the notable exception that the photon number is not directly
affected,
there are many similarities with damping.

For simulation, parameters
 are chosen so that $\epsilon
g/\Gamma =\pi /2$, $\lambda =2.2$ and $\Gamma/2\epsilon^{2} =0.001$.  The last
equality maintains the phase diffusion term at a small and
constant
level. This ensures that the same state
 is always
 produced by the adiabatically
eliminated master equation.

It is worth mentioning how the full dynamics were simulated.
Due to the jump in the field amplitude of the ancilla cavity when a
detection on the output of the system is made, the basis size required for
an accurate simulation is large.  If the amplitude jumps by an amount
$\epsilon /2$ then the photon number will increase by (presuming the
initial field was small) $\epsilon^{2}/4$.
A second detection on the system occurring very soon after the
first, would push the photon number even higher.
In fact the computational resources
available were not sufficient to allow even a quantum trajectory
simulation \cite{opsystem,wfuncandqjump,qmonte} of \erf{eqnMEtwocav}.
The solution was to make a unitary
transformation to a frame in which the evolution of the driven cavity
due to feedback was separated from that due to quantum noise.  That is,
the mean amplitude of the field was described classically while the quantum
representation of the noise was maintained.  The unitary transformation
used was
\beq
U=\exp[\epsilon f(t)(b-b^{\dag})/2],
\label{unit1}
\eeq
where $f(t)$ is defined by
\beq
f(t)=\int_{-\infty}^{t}ds\exp[-\Gamma (t-s)/2]I(s).
\label{ftee}
\eeq
Here, $I(t)$ is the $c$-number stochastic photocurrent.
The price of a reduced basis size is a time dependent Liouvillian.  When
the transformation of \erf{unit1} is applied to the implicit master
equation (feedback is described by a feedback Hamiltonian instead of
the exponentials) an
equation is obtained that is already of an explicit form (see
appendix~\ref{elbow})
\bqa
\dot{W}&=&-i\left[ga^{\dag}a\{b+b^{\dag}+\epsilon
f(t)\}-\frac{i\lambda}{4}\left\{a^{2}-(a^{\dag})^{2}\right\},W\right]\nn\\
&&+{\cal D}[a]W+\Gamma {\cal D}[b]W.
\eqa
It can be seen that $\epsilon f(t)$ represents the amplitude of the driven
cavity.  Although $f(t)$ is stochastic, it is a smoothed
(non-singular) version
of the photocurrent  and can therefore be treated without worrying about the
stochastic calculus. Note also that since $U$ contains only ancilla
operators, the system state matrix $\rho={\rm Tr}_{b}[\tilde{W}]$ is
the same as before, ${\rm Tr}_{b}[W]$.

The transformed master equation was simulated using quantum trajectory
methods.  It is shown in Fig.~\ref{eocavsubplot} that as $\Gamma$
 becomes large the
adiabatically eliminated master equation becomes a very good approximation
to the
full dynamics.  Clearly, though, $\Gamma$ has
to be pushed to much higher levels than  the TLA damping for this
correspondence to hold.  One reason for this is that the Wigner functions of
the steady state density matrices for electro-optic feedback onto a mode have
much greater structure, meaning that a measure such as the Bures distance
(which measures the distinguishability of states) will be more sensitive
to small differences.  It also is likely that the parameter regime chosen
is one in which this system varies quickly, with the result that adiabatic
elimination will only be valid at very large $\Gamma$.

A comparison of the Wigner functions [Figs.~\ref{eocavsubplot}(a) and (c)]
with that produced with simple feedback [Fig.~\ref{sfbsubplot}(a)] shows
the expected similarity.  It is expected because \erf{eqnMEadtwocav} only
 differs
from simple feedback due to the presence of the double commutator noise
term, which was chosen to be small.

\subsection{All-optical Feedback onto an Atom}

The basis size of the TLA ensures that simulating the full dynamics
of all-optical feedback
[\erf{eqnMEallopt}] is relatively easy.  However, a threshold driving
strength exists ($\lambda
=1$) for this system which means that the adiabatically
eliminated master equation cannot be tested
in the same regime as the previous sections.  Instead we set
$\lambda =0.97$ which enabled us to perform an accurate simulation
with the computational resources available.

The Hamiltonian of \erf{eqnMEallopt} includes the parametric
amplifier driving and also the coupling of \erf{qed}.  Once again we choose
$K=ga^{\dag}a$ and set $4g/\Gamma =\pi /2$, while varying $\Gamma$ and $g$.
The
Bures distance between the states produced by \erf{eqnMEallopt} and
\erf{eqnadallopt} is
shown in Fig.~\ref{alloptatsubplot}, as are some Wigner functions
for the full dynamics
and the adiabatic state.  It can be seen that the state produced with
the full dynamics
approaches the adiabatic state at a similar rate,
as $\Gamma$ is increased, to
electro-optic feedback via a TLA.

There is a large similarity between the state
produced via simple feedback
 in Fig.~\ref{sfbsubplot} (c) and that in Fig.~\ref{alloptatsubplot} (a), with
 the presence of shearing being the most
notable difference.  This closer correspondence to simple feedback
than that of the electro-optic feedback systems is not surprising given that
the adiabatic all-optical master equation was the same as simple feedback
to a higher order (second).  The
smaller driving also contributes to the closeness of the states.

\subsection{All-optical Feedback via a Mode}
It was shown in Sec.~\ref{ee} that in the adiabatic limit
all-optical feedback onto a mode has the same effect as feeding back
onto a TLA.  Therefore, the same threshold for the driving strength
exists for this system ($\lambda =1$).

 The basis size required here is not as large as for electro-optic
feedback because the photons leak out of the system and into the
ancilla
cavity, giving a smooth variation of photon number.  Despite this, a
quantum trajectory simulation was still found to be necessary.  The
 results obtained for $\lambda =0.97$ and $4g/\Gamma =\pi /2$ can be found
in Fig.~\ref{alloptcavsubplot}.  The adiabatic state is, of course,
 the same as for all-optical feedback onto a
TLA.  There is a notable difference in the speed at which the full dynamics
 approaches this state.  At low damping the Bures distance is already very
 low.  The conclusion is that the ancilla mode has minimal effect on the
  system when included in an all-optical feedback loop.

\subsection{Comparison with ``Reversible Feedback'' Generated by a
$\chi^{(3)}$ Non-linearity}
\label{nonlin}
Finally, we consider the effect of placing a $\chi^{(3)}$
material inside an optical cavity driven by a parametric oscillator.
There is no feedback loop involved.
The Hamiltonian generated by the $\chi^{(3)}$ non-linearity (a Kerr
non-linearity) is given by \cite{Drumkerr}
\beq
H_{{\rm Kerr}}=\frac{\chi}{2}(a^{\dag})^{2}a^{2}.
\label{Kerr}
\eeq
The Heisenberg equation of motion of the annihilation operator
due to this Hamiltonian is found to be
\beq
\dot{a}=-i\chi (a^{\dag}a)a.
\eeq
Thus, it is clear that the $\chi^{(3)}$ non-linearity causes a
detuning proportional to the intensity of the field inside the cavity.  In
this
way the system has a self-awareness that is similar to simple
feedback, which is why a comparison is relevant.  In fact, it can be
shown that the two systems are classically equivalent given the same
choice of the parameter $\chi$.
For large feedback the two systems diverge when treated quantum
mechanically.  One of the main reasons behind this is that the Kerr
effect displays no periodic dependence upon its magnitude, whereas
the simple feedback does.  This is illustrated in
Fig.~\ref{chisubplot}(c), where the
Bures distance between the steady states of the two systems is plotted
for varying $\chi$.  The Wigner function of
the ``reversible feedback" steady state for $\chi =\pi /2$ with $\lambda
=2.2$ and $\lambda =0.97$ is given in Figs.~\ref{chisubplot}(a) and (b)
respectively.  They are seen to be very different from any of the
steady states produced by feedback.

\section{Discussion}

\subsection{Summary}
The description of feedback in compound quantum systems (where
the output
from the system is used  to control the
evolution of the ancilla, which is reversibly coupled to the system)
is greatly simplified if the ancilla can be adiabatically eliminated.
We have shown how this can be done for four generic cases, arising
from considering two forms of feedback (all optical or coherent, and
electro-optical or incoherent) and
two types of ancilla (a two-level atom, and an optical mode).
The four resulting master equations for the system alone are given below.
They are  the most
important results of this paper. We also include the perturbative
expansions of these master equations to third order in the feedback
operator $Z$. All of the equations are identical to first order in
$Z$, but differ in second or third order.

For comparison, we begin with  simple feedback (that is, with no
ancilla) based on detection of the intensity $I=b_{\rm out}\dg
b_{\rm out}$ of the output field $b_{\rm out}=b_{\rm in}+c$,
and using the feedback Hamiltonian
\beq
H_{\rm fb} = I(t)Z.
\eeq
The master equation for this is
\bqa \label{dirfbme}
\dot{\rho} &=& -i[H,\rho]+{\cal D}[e^{-iZ}c]\rho \\
 &\simeq &-i[H,\rho  ]+{\cal D}[c]\rho + {\cal C}[Z]{\cal J}[c]\rho \nn\\
&&+\;\cu{\frac{1}{2}\left({\cal
C}[Z]\right)^{2}+\frac{1}{6}\left({\cal C}[Z]\right)^{3}}
{\cal J}[c]\rho  .
\eqa
Here ${\cal C}[Z]B \equiv -i[Z,B]$ as before. The remaining master equations
result from trying to reproduce this form of feedback via an ancilla.

The first master equation
derived using adiabatic elimination  is for electro-optic feedback via
the inversion of a
two-level atom:
\bqa
\dot{\rho} &=&-i[H ,\rho ]+\int_{0}^{\infty}dq e^{-q}{\cal
D}[e^{-iqZ}c]\rho , \label{fmeduae} \\
&\simeq&-i[H,\rho] + {\cal D}[c]\rho+ {\cal C}[Z]{\cal J}[c]\rho
\nl{+}\cu{({\cal C}[Z])^{2}+({\cal C}[Z])^{3}}{\cal J}[c]\rho.
\eqa
This differs from the simple feedback master equation (\ref{dirfbme})
at second order
in $Z$.
The second is for electro-optic feedback via one quadrature of an optical mode:
\bqa \label{eefq}
\dot{\rho} &=&-i[H ,\rho] + {\cal D}[e^{-iZ}c]\rho +\frac{\Gamma}{\epsilon^{2}}
{\cal D}[Z].
\eqa
The expansion of the above equation can be found from that of the simple
feedback. The size of the extra second-order term is determined by
$\Gamma$, the damping rate for the ancilla mode, and $\epsilon$,
the strength of driving of the ancilla mode.

Turning now to all-optical feedback, we have found that the same
master equation arises regardless of whether the feedback is via
the inversion of two-level atom or the intensity of an optical mode. It is:
\bqa
\dot{\rho} &=&-i[H
,\rho]+{\cal D}\left[\exp
\left(-2i\arctan\frac{Z}{2}\right)c\right]\rho, \\
&\simeq &-i[H,\rho  ]+{\cal D}[c]\rho+ {\cal C}[Z]{\cal J}[c]\rho  \nn\\
&&+\cu{\frac{1}{2}\left({\cal
C}[Z]\right)^{2}+\frac{1}{4}{\cal C}[Z]({\cal J}[Z]
-2{\cal A}[Z]) }{\cal J}[c]\rho .\nn\\
\eqa
Unlike \erf{fmeduae}, this differs from \erf{dirfbme} only at third
order in $Z$.

\subsection{Relation to Previous Work}
As mentioned in the introduction Slosser and Milburn \cite{SloMil94} perform
adiabatic elimination of the pump mode of a non-degenerate parametric
oscillator.  In their system the pump mode is driven by the output
photocurrent from the idler mode.  The procedure they adopt is similar
to that contained in Sec.~\ref{BB} of this paper, in that they expand
the density matrix in terms of the lower number states of the pump mode.
 However, in Sec.~\ref{elecmode} we have already noted that this is not
 appropriate when dealing with direct detection feedback onto a {\em mode}.
 Higher number states are essential to
  the description of the system if the feedback strength is large.
   For this reason they limit the feedback strength to small and
   moderate values, with a generalization to larger feedback contained
   in their appendix.  This appendix does not explain the origin of
   the all-orders feedback term.  The techniques of adiabatic elimination
    using QLE's that are presented in this paper make it easy
    to treat their system rigorously to all-orders in the feedback strength.
    The final result, using their definitions and our superoperators,
    is (with perfect detection assumed):
\bqa
\dot{\rho}&=&\epsilon [a^{\dag}b^{\dag}-ab,\rho ]+2\Gamma{\cal D}[ab]\rho
+\gamma_{a}{\cal D}[a]\rho\nn\\
&&+\gamma_{b}{\cal D}[\exp\ro{\chi ab-\chi a^{\dag}b^{\dag}}b]\rho .
\eqa
Note that the second term here is the one analogous to the final term
in our \erf{eefq}.

Doherty and co-workers consider a strongly interacting system comprised of
an atom inside a cavity \cite{DohParTanWal98}.
The methods used for adiabatic elimination
are similar to those used in this paper.
They form QLE's for operators from any of the three subsystems (center of mass
motion, internal state and the cavity mode) and then set the time derivatives
of, first, the internal state operator and, second, the cavity operator, to
zero.
 They then substitute into the QLE for the momentum operator
 $p_{x}$.
After a conversion to the explicit form of the QLE,
they show that the QLE they derive is compatible with  the master
equation (using their notation)
\beq
\dot{\rho} = -\frac{i}{\hbar}\sq{\frac{p_{x}^{2}}{2m},\rho}
+ \frac{\kappa }{2} {\cal D}\sq{\exp\ro{
-2i\arctan\frac{Z}{2}}\alpha}\rho,
\eeq
where
\beq
Z = \frac{g_{0}^{2}\cos^{2}k_{L}x}{\Delta \kappa}.
\eeq
Note the similarity with our equation resulting from adiabatic
elimination of an optical mode where the coupling is via the
intensity, (but of course there is no feedback here so our operator $c$
is replaced by the $c$-number $\alpha$). The derivation of this master equation
in Ref.~\cite{DohParTanWal98} is not completely rigorous in that
other master equations would also be compatible with the QLE they
derive for $p_{x}$. However, it would be straightforward, using the
technique we introduced in Sec.~\ref{ee}, to make it rigorous.

The work done on all-optical feedback in this paper follows on
from that done by Wiseman and Milburn \cite{WisMil94b}.  They were able
to show that all-optical feedback onto a mode could replicate
electro-optic homodyne-detection feedback, but they could only prove
equivalence with
direct-detection feedback to second order.  Here, we have shown
that this is because the equivalence only holds to second order.
We have done this by finding the master
equation to all-orders in the feedback strength, and showing it to be
of the Lindblad
form.

Showing that all-optical feedback via an ancilla (be it a
two-level atom or a mode) cannot replicate electro-optical direct
detection feedback, leads naturally to the question of whether a
more complicated
all-optical feedback scheme can replicate direct electro-optic feedback.
Since the feedback is replicated to second order, a fruitful approach
would seem to be to make the feedback weak, while multiplying up the
number of ancillae to compensate. In appendix~\ref{inf} we consider
the case of $N$   ancillae, with coupling to the system scaling as
$1/N$, where the output of the system is fed sequentially into all of
the ancillae. We show that in the limit $N\to \infty$, this
hypothetical all-optical feedback scheme does indeed produce
the simple electro-optic feedback master equation (\ref{dirfbme}).

\subsection{Conclusion}
We have shown that it is possible to greatly simplify the description
of feedback in compound quantum systems by adiabatically eliminating
the ancilla, to give master equations for the system alone.
 In essence, we have found the first order in $\Gamma^{-1}$ effect of
 the ancilla upon the system, where $\Gamma$ is the ancilla decay rate.
We have done this for a variety of ancillae and forms of feedback,
and found good agreement with  numerical simulations of the
dynamics for the full compound quantum system. The master equations
in the various cases are quite different, and their range of validity
(that is, how large $\Gamma$ has to be for them to be valid) was also
found numerically to differ. For the numerical simulations we
of course used a particular system, but the equations we derive are
very general.

  The primary motivation for this work
is the reduction of basis size that is necessary to describe the evolution
 of the system.  It is hoped that the derived equations will prove
  to be helpful to co-workers. However, we note that
  numerical testing (to find the regime
  in which these equations are a good
  approximation) may be necessary to determine when it is appropriate
  to use them.  Apart
  from these practical advances, we feel that the previously existing
  confusion in the literature, as discussed in the introduction, has
  been resolved, and the procedure of
  adiabatic elimination in compound quantum  systems with feedback
  is now on stable ground.

\acknowledgments
We would like to acknowledge discussions with W.J. Munro and S.M. Tan.
This work was supported in part by the Australian Research Council.

\appendix

\section {Proof of Lindblad Form}
\subsection{Electro-optic Feedback onto a TLA}
\label{APPlindtla}
To show that \erf{hello} can be written in the Lindblad form the
following identity will first be established:
\begin{equation}
(\Gamma -{\cal C}[K])\int_{0}^{\infty}dx{\cal J}\left[e^{
-x(\Gamma +2iK)/2}\right]\rho=\rho .
\label{weirdthing}
\end{equation}
Multiplying the equation through by two arbitrary eigenstates of $K$,
$\langle \alpha |$ and $|\beta \rangle$, from the left and right
respectively, the following is obtained:
\begin{equation}
\rho_{\alpha\beta}[\Gamma+i(\alpha -\beta
)]\int_{0}^{\infty}dx e^{-x[\Gamma+i(\alpha -\beta )]}
=\rho_{\alpha\beta }.
\end{equation}
After the simple integration is performed the identity is proved.
Before using this the following rearrangement is made:
\beq {\cal C}[Z](1 -{\cal C}[Z])^{-1}=(1 -{\cal C}[Z])^{-1}-1 .
\label{superop}
\eeq
Upon use of the identity with $Z=K/\Gamma$
the master equation \erf{hello} becomes \erf{berry}.

\subsection{All-optical Feedback onto an Atom}
\label{aloptlind}
In order to show that the master equation can be written as in
\erf{g'daymate} it is sufficient to show that
\bqa
&\exp \left[-2i\arctan\left({Z}/{2}\right)\right]\rho\exp
\left[2i\arctan\left({Z}/{2}\right)\right]-\rho\nn\\
& ={\cal C}[Z]{\cal
J}\left[\left(1+i{Z}/{2}\right)^{-1}\right]\rho .
\label{eqnexp}
\eqa
Note that ${\cal J}[c]$ has been omitted as it is a multiplicative
factor on both of the superoperators.  Consider the following
non-Hermitian
operator that can be put into a modulus and argument form:
\begin{equation}
\frac{1+ {iZ}/{2}}{\sqrt{1+\left({Z}/{2}\right)^{2}}}=re^{i\theta}.
\label{eqneitheta}
\end{equation}
That $r=1$ can be quickly verified.  The argument is given by
\begin{equation}
\theta ={\rm arctan}\left({Z}/{2}\right).
\end{equation}
By taking the logarithm of~\ref{eqneitheta} an alternative expression
for the argument is obtained
\begin{equation}
\theta =\ln
\left(\frac{\sqrt{1+\left({Z}/{2}\right)^{2}}}{1+{iZ}/{2}}\right)^{i}.
\end{equation}
If the logarithmic form of ${\rm arctan}\left(Z/2\right)$ is
used, then the exponentials of \erf{eqnexp} disappear.  The
LHS of that equation becomes
\begin{equation}
\frac{1+\left(Z/2\right)^{2}}{\left(1+ iZ/2\right)^{2}}
\rho\frac{\left(1+ iZ/2\right)^{2}}{1+\left(Z/2\right)^{2}}-
\rho .
\end{equation}
It is now noted that
$1+\left(Z/2\right)^{2}=\left(1+ iZ/2\right)\left
(1- iZ/2\right)$.
The above expression can be re-written as
\begin{eqnarray}
\frac{1- iZ/2}{1+ iZ/2}\rho\frac{1+ iZ/2}{1- iZ/2}
-\frac{1+ iZ/2}{1+ iZ/2}\rho\frac{1- iZ/2}{1-iZ/2}.
\eqa
After algebraic manipulation this can be shown to be equal to the RHS
of \ref{eqnexp}, as required.

\section{Electro-optic Feedback via an atom with Jaynes-Cummings
coupling and detuning}
\label{JC}

In this section we take the compound system as being a single mode
optical cavity, with electro-optic feedback onto a
TLA that is placed in the standing wave of the cavity.
The Jaynes-Cummings coupling that will be used is
$V=g(a\sigma^{\dag}+\sigma
a^{\dag})$, with $g$ being a real constant and $a$ the
annihilation operator for the cavity mode.  A detuning of
$\delta\sigma^{\dag}\sigma$ is also included.  The following hierachy
of parameters will be investigated:
\beq
\delta\gg g\gg\Gamma\gg{\cal C}[H_{\rm s}].
\eeq
Of course, ${\cal C}[H_{\rm s}]$ is really a superoperator (containing
the system Hamiltonian terms)
so here we are only
referring to its scalar part.

As will be shown, when the the necessary variables are slaved
a Hamiltonian term of the form
$g^{2}a^{\dag}a/\delta$ is obtained in the final master equation.
With the above scaling, this Hamiltonian is not necessarily small
compared to $\Gamma$. This makes the adiabatic elimination
of the atom   more difficult since the presumption that the atomic
relaxation time is much shorter than any system time scale is not
necessarily true.  To do
the elimination of the atom rigorously we therefore transform to an
interaction
picture defined by
$H_{0}=-g^{2}(a^{\dag}a+\sigma^{\dag}\sigma)/\delta$.  This
transformation has the additional effect of adding a time dependence
into the feedback term of the master equation.  To nullify this we
will start with a time dependent feedback Hamiltonian whose effect, when
moved to the interaction picture, is time independent.  The master
equation in the \sch picture is thus
\bqa
\dot{W}&=&{\cal C}[H_{\rm s}]W-i[\delta\sigma^{\dag}\sigma
+g(a\sigma^{\dag}+\sigma a^{\dag}),W]+\Gamma {\cal
D}[\sigma ]W  \nn\\
&&+{\cal D}[\cu{\sigma\exp(-ig^{2}t/\delta )
+\sigma^{\dag}\exp(ig^{2}t/\delta )}a]W.
\eqa

In the
interaction picture with respect to $H_{0}$ the master equation is
\bqa
\dot{\tilde{W}}&=&U^{\dag}{\cal C}[H_{\rm s}](U\tilde{W}U^{\dag})U-i\left[g^{2}(a^{\dag}a+
\sigma^{\dag}\sigma
)/\delta ,\tilde{W}\right]\nn\\
&&-i[g(a\sigma^{\dag}+\sigma
a^{\dag})+\delta\sigma^{\dag}\sigma,\tilde{W}]+\Gamma{\cal
D}[\sigma ]\tilde{W}\nn\\
&&+{\cal
D}[\exp(-i\pi\sigma_{x}/2)a]\tilde{W}.
\eqa
For simplification we will put $
\Delta =\delta +g^{2}/\delta$.

The expansion of \erf{rhoexpan} is made, with the $\rho$s
now understood to be in the interaction picture.  The time rates of
change are
\begin{eqnarray}
\dot{\rho_{0}}&=&{\cal C}[\tilde{H}_{\rm s}]\rho_{0}+{\cal
J}[a]\rho_{2}-{\cal A}[a]\rho_{0}+\Gamma\rho_{2}\nn\\
&&-ig(a^{\dag}\rho_{1}^{\dag}-\rho_{1}a)-
i\left[\frac{g^{2}a^{\dag}a}{\delta},\rho_{0}\right],\\
\dot{\rho_{1}}&=&{\cal C}[\tilde{H}_{\rm s}]\rho_{1}+{\cal
J}[a]\rho_{1}^{\dag}-{\cal
A}[a]\rho_{1}+ig(\rho_{0}a^{\dag}-a^{\dag}\rho_{2})\nn\\
&&-\left(\frac{\gamma}{2}-i\Delta\right)\rho_{1}-i\left[\frac{g^{2}a^{\dag}a}
{\delta},\rho_{1}\right],\\
\dot{\rho_{2}}&=&{\cal C}[\tilde{H}_{\rm s}]\rho_{2}+{\cal
J}[a]\rho_{0}-{\cal A}[a]\rho_{2}-\Gamma\rho_{2}\nn\\
&&-ig(a\rho_{1}-\rho_{1}^{\dag}a^{\dag})-i\left
[\frac{g^{2}a^{\dag}a}{\delta},\rho_{2}\right].
\label{eqnrho2dot1}
\end{eqnarray}
When $\rho =\rho_{0}+\rho_{2}$ is used, the above equations give
\bqa
\dot{\rho} &=& {\cal C}[\tilde{H}_{\rm s}]\rho+{\cal
D}[a]\rho-i\left[g^{2}a^{\dag}a/\delta ,\rho\right]
\nl{-}ig [a^{\dag},\rho_{1}^{\dag}]-ig[a,\rho_{1}].
\label{eqndotrhocav1}
\eqa

In the limit $\Gamma\gg{\cal C}[\tilde{H}_{\rm s}]$ the amplitudes
of $\rho_{1}$ and $\rho_{2}$ respond to changes in the cavity mode much more
quickly than $\rho_{0}$.  Their equilibrium values are
\bqa
\rho_{1}&=&\left(\Delta-\frac{\Gamma}{2i}\right)^{-1}
\left(1-\frac{ig^{2}}{\Delta\delta}\left(1-\frac{\Gamma}{2i\Delta}\right)^{-1}
{\cal C}[a^{\dag}a]\right)^{-1}\nn\\
&&\times (-g\rho a^{\dag}
+g^{2}\{a^{\dag},\rho_{2}\})\label{helloo},\\
\rho_{2}&=&\frac{1}{\Gamma}\left(1+\frac{{\cal J}[a]}{\Gamma}-
\frac{g^{2}}{\Gamma\delta}{\cal C}[a^{\dag}a]\right)^{-1}\nn\\
&&\times \cu{ {\cal J}[a]\rho
-ig(a\rho_{1}-\rho_{1}^{\dag}a^{\dag})}.
\eqa
These two equations can be rearranged to give $\rho_{2}$ in terms of
$\rho_{0}$.  In the limit $\Delta\approx\Gamma^{2}$ and
$g^{2}\approx\Gamma\Delta$, we find to first order
\beq
\rho_{2}=\frac{1}{\Gamma}\left(1-\frac{2g^{2}}{\Delta\Gamma}{\cal
C}[a^{\dag}a]\right)^{-1}
{\cal J}[a]\rho
+\frac{g^{2}}{\Delta^{2}}{\cal J}[a]\rho .
\label{rhotoo}
\eeq
Using this in \erf{helloo} allows
the following master equation to be derived:
\bqa
\dot{\rho}&=&{\cal C}[\tilde{H}_{\rm s}]\rho+{\cal D}[a]\rho+
\frac{\Gamma g^{2}}{\Delta^{2}}{\cal D}[a]\rho
 -i\left[\frac{g^{4}(a^{\dag})^{2}a^{2}}{\Delta^{3}},\rho\right] \nn\\
&&+\frac{2g^{2}}{\Delta\Gamma}{\cal C}[a^{\dag}a]\left(1-\frac{2g^{2}}
{\Delta\Gamma}
{\cal C}[a^{\dag}a]\right)^{-1}{\cal J}[a]\rho .
 \eqa
In the limit of $\Delta\gg\Gamma^{2}$, while still maintaining
$g^{2}\approx\Gamma\Delta$, the third and fourth terms drop
out, leaving the same master
equation derived in Sec.~\ref{BB},
with $Z=2g^{2}a^{\dag}a/\Gamma\Delta$ of order unity.
This is the same limit in which Walls and Milburn arrive
at the effective Hamiltonian used in \erf{sigma+1} \cite{wallsmilburn}.
The third and fourth terms correspond to, respectively,
 an increased damping rate and a
$\chi^{(3)}$ nonlinearity for the cavity mode.

Note that the derived Hamiltonian term which
threw doubt upon the adiabatic elimination process has been canceled.  Of
course, when we return to the \sch picture it will reappear,
leaving a different master equation from that of Sec.~\ref{BB}.  The
solution is to start with an extra Hamiltonian term of the form
$-g^{2}a^{\dag}a/\Delta$ when using the Jaynes-Cummings coupling.
A transformation to the interaction picture is then not required, nor
is the time dependence in the feedback.

\section{Unitary Transformation of the Total Master Equation for
Electro-optic Feedback Onto a Mode}
\label{elbow}
In this appendix the total master equation for electro-optic feedback
onto a cavity is unitarily transformed so that the amplitude of the
driven cavity may be treated classically, thus reducing the necessary
basis size.
The implicit form of the master equation will be used as this proves
to be more straightforward.  That is to say, the photocurrent will be
approximated by a slightly smoothed version of $dN/dt$.
Before transformation the implicit master equation is
\bqa
\dot{W}&=&-i\left[g
a^{\dag}a(b+b^{\dag})+\frac{i\lambda}{4}\{(a^{\dag})^{2}-a^{2}\},W\right] 
+{\cal D}[a]W\nn\\
&&-\frac{i\epsilon}{2}\left[
(-ib+ib^{\dag})I(t),W\right]+\Gamma{\cal D}[b]W.
\label{unit}
\eqa
We now put $\tilde{W}=UWU^{\dag}$ where,
\beq
U=\exp[\epsilon f(t)(b-b^{\dag})/2]
\eeq
and $f(t)$ is given in \erf{ftee}.  The unitarily transformed master
equation is given by
\beq
\dot{\tilde{W}}=\dot{U}U^{\dag}\tilde{W}+
\tilde{W}U\dot{U}^{\dag}+U{\cal L}(U^{\dag}\tilde{W}U)U^{\dag},
\label{youdot}
\eeq
where $\dot{W}={\cal L}W$.  Now, $\dot{U}$ is given by
\beq
\dot{U}=\frac{\epsilon}{2} (b-b^{\dag})[-{\Gamma f(t)}/{2}+I(t)]U,
\eeq
thus the first two terms of \erf{youdot} give
\begin{equation}
\frac{\epsilon}{2} \left[-\Gamma f(t)/2+
I(t)\right][b-b^{\dag},\tilde{W}].
\label{who}
\end{equation}
The last term of \erf{youdot} will only cause a change to
terms that are dependent upon the driven cavity operators.  Thus, the
non-linear driving and the damping of the system may be ignored for
the present.  The expression that needs to be simplified contains
three terms (damping, coupling, and feedback),
which can be evaluated using $UbU^{\dag}={\epsilon}
f(t)/2+b$. The damping term is
\begin{eqnarray}
&&\Gamma U\cu{{\cal D}[b](U^{\dag}\tilde{W}U)}U^{\dag}\nn\\
&&=\Gamma \left({\cal D}[b]\tilde{W}+{\epsilon
f}\left[b-b^{\dag},\tilde{W}\right]/4\right),
\eqa
the coupling term is
\bqa
&&-iU[ga^{\dag}a(b+b^{\dag}),U^{\dag}\tilde{W}U]U^{\dag}\nn\\
&&=-i[ga^{\dag}a(b+b^{\dag}+\epsilon
f),\tilde{W}],
\eqa
and the feedback term is
\bqa
&&-\frac{iU}{2}\left[\epsilon
(-ib+ib^{\dag})I(t),U^{\dag}\tilde{W}U\right]U^{\dag}\nn\\
&&=-\frac{\epsilon}{2} \left[I(t)(b-b^{\dag}),\tilde{W}\right].
\eqa
Adding up the contributions from the damping, feedback, coupling,
 \erf{who} and also the system Hamiltonian, the following is
obtained:
\bqa
\dot{\tilde{W}}&=&-i\left[ga^{\dag}a(b+b^{\dag}+\epsilon
f)+\frac{i\lambda}{4}\{(a^{\dag})^{2}-a^{2}\},\tilde{W}\right]\nn\\
&&+\Gamma{\cal
D}[b]\tilde{W},
\eqa
which is the master equation after transformation.  There is now no
distinction between the implicit and explicit forms as $f$ is a
bounded function.

\section{All-optical Feedback onto an Infinite Number of Cavities}
\label{inf}
It is of interest whether all-optical feedback can ever have the same
effect on a  system as simple electro-optic feedback.  In this
section we show that this can be achieved with a very large
number of ancilla cavities that are coupled back to the system.
The basic idea
of
the all-optical feedback remains the same, in that the output of one
cavity becomes the input to the next cavity.  The cavities all have the same
damping
co-efficient, $\Gamma$.  Damping of the system is
set
equal to unity.  For large damping, the infinite number of ancilla
cavities
will be adiabatically eliminated.  See
Fig.~\ref{alloptinf2}.

The form of the coupling of the $j^{{\rm th}}$ cavity
is similar to that for the
all-optical
feedback via a single mode.  It is
\begin{equation}
V_{j}=\frac{Kb^{\dag}_{j}b_{j}}{N},
\label{hect}
\end{equation}
where $N$ is the total number of driven cavities, $K$ is
proportional to
an Hermitian system operator and $b_{j}$ is the annihilation operator
of
the $j^{{\rm th}}$ cavity.  The input and output fields to and from
the $j^{{\rm th}}$
cavity are represented by $u_{{\rm in}(j)}$ and $u_{{\rm out}(j)}$.
This means that $u_{{\rm in}(j+1)}=u_{{\rm out}(j)}$.  The output field
from the system is given by $v_{{\rm
out}}$.

From \erf{eqndsallopt} the contribution
to
the QLE for an arbitrary operator due to the shining of the
$(j-1)^{{\rm th}}$
output field onto the $j^{{\rm th}}$ cavity can be found.  The total QLE if
there
are $N$ driven cavities is
\begin{eqnarray}
dr&=&{\cal D}[c^{\dag}]rdt-
[dV^{\dag}_{{\rm
in}}c-c^{\dag}dV_{{\rm in}},r]
+\Gamma\sum^{N}_{j=1}{\cal D}[b^{\dag}_{j}]rdt\nn\\
&&-\sqrt{\Gamma}\sum^{N}_{j=1}[dU^{\dag}_{{\rm
in}(j)}b_{j}-b_{j}^{\dag}dU_{{\rm
in}(j)},r]\nonumber\\
&&+i[H_{{\rm s}},r]dt+i\sum^{N}_{j=1}b^{\dag}_{j} [\frac{K}{N},r]
b_{j}dt,
\end{eqnarray}
where $dU_{{\rm in}(j)}=u_{{\rm in}(j)}dt$.
Note that the same idea as in \erf{trick} has been used.  The QLE
for a
system operator is
\begin{eqnarray}
ds&=&{\cal D}[c^{\dag}]sdt-[dV^{\dag}_{{\rm
in}}c-c^{\dag}dV_{{\rm in}},s]\nonumber\\
&&+i\sum^{N}_{j=1}b^{\dag}_{j}[\frac{K}{N},s]b_{j}dt+i[H_{{\rm s}},s]dt.
\label{c}
\end{eqnarray}
Although the input fields $u_{{\rm in}(j)}$ obviously depend on
system
operators, they are evaluated at a slightly earlier time due to the
small,
but finite, propagation time of the field from the system to the
driven
cavities.  The system operator $s$, therefore, commutes with
$dU_{j}$.

We now note that the QLE for $b_{j}$ has the same form as
\erf{eqndotballopt}
\beq
db_{j}=-\left(\frac{\Gamma b_{j}}{2}+\sqrt{\Gamma}u_{{\rm
in}(j)}+\frac{iKb_{j}}{N}\right)dt.
\label{a}
\eeq
Also,
\beq
u_{{\rm in}(j+1)}=u_{{\rm in}(j)}+\sqrt{\Gamma}b_{j}.
\label{b}
\eeq
To simplify matters only the non-stochastic part of $b_{j}$ and
$u_{{\rm
in}(j)}$ will be considered in the derivation of the master equation
for
the system.  This can be justified by mathematical induction.
Suppose
that $u_{{\rm in}(j)}$ and $b_{j}$ can both be grouped into
stochastic
terms linearly dependent upon $v_{{\rm in}}$ and non-stochastic
terms.
Then it is clear from \erf{b} that $u_{{\rm in}(j+1)}$ can
also be
grouped in such a manner.  Therefore, in the limit in which
\erf{a} can be slaved to produce the equivalent of \erf{retro} it
can be
seen that $b_{j+1}$ will consist of non-stochastic terms, arising
from the
non-stochastic terms of $u_{{\rm in}(j)}$ and $b_{j}$, as well as
stochastic terms linear in $v_{{\rm in}}$.  To complete the
mathematical
induction, $b_{1}$ and $u_{{\rm in}(1)}$ can obviously grouped in the
manner suggested.  Now, terms in $b_{j}$ that go as $v_{{\rm in}}$
will
annihilate onto the vacuum state when the expectation value of
\erf{c} is taken, thus, the stochastic parts can be ignored
as
they give a zero contribution.

An expression for non-stochastic part of $u_{{\rm in}(j)}$ (denoted
by
$\bar{u}_{{\rm in}(j)}$) needs to be found in order to evaluate
$\bar{b}_{j}$. Using \erf{b} and the slaved value
$\bar{b}_{j-1}$ it is found to be
\beq
\bar{u}_{{\rm in}(j)}=\left(\frac{-1+2iK/\Gamma N}{1+2iK/
\Gamma N}\right)^{j-1}c.
\eeq
Substituting into \erf{a} gives $b_{j}$.  This
is
then used in \erf{c}.  Writing $Z=4K/\Gamma$,
the summation term is
\bqa
&&\frac{i}{N}{\cal
J}\left[c^{\dag}\left(1-iZ/2N\right)^{-1}\right]\nn\\
&&\times\sum^{N}_{j=0}\left(\frac{-1-iZ/2N}{1-iZ/2N}\right)^{j}
[Z,s]\left(\frac{-1+iZ/2N}{1+iZ/2N}\right)^{j}.
\eqa
Firstly, the quotients are expanded
to
second order in $Z/N$.  Then the contributions from the first
and second orders are factorized, with the latter expanded to first order in
$j/N^{2}$.  This gives
\bqa
&&\frac{i}{N}{\cal
J}\left[c^{\dag}\left(1-iZ/2N\right)^{-1}\right]\nn\\
&&\times\sum^{N}_{j=0}\left(1+iZ/N\right)^{j}\left(1-\frac{j}{2}
\left(Z/N\right)^{2}\right)[Z,s]\nn\\
&&\times\left(1-iZ/N\right)^{j}\left(1-\frac{j}{2}
\left(Z/N\right)^{2}\right).
\label{summb}
\eqa
It is not difficult to show that the contribution of the $j/N^{2}$ terms
is small (of order $N^{-1}$).  Also, $(1+iZ/2N)^{-1}\simeq 1$,
so we can now write the summation as
\bqa
&&i\sum^{N}_{j=0}{\cal
J}\left[c^{\dag}\left(1+iZ/N\right)^{j}\right]\left[Z/N,s\right
]\nn\\
&\simeq &\sum^{N}_{j=0}{\cal
J}\left[c^{\dag}\left(1+iZ/N\right)^{j}\right]\left({\cal J}
\left[\left(1+iZ/N\right)\right]-1\right)s\nn\\
&=&{\cal J}[c^{\dag}]\sum^{N}_{j=0}\left({\cal
J}\left[\left(1+iZ/N\right)^{j+1}\right]-{\cal
J}\left[\left(1+iZ/N\right)^{j}\right]\right)s\nn\\
&\simeq &{\cal J}[c^{\dag}]\left({\cal
J}\left[\left(1+iZ/N\right)^{N}\right]-1\right)s\nn\\
&\rightarrow &{\cal J}[c^{\dag}]({\cal J}[\exp(iZ)]-1)s.
\eqa
Terms of order $N^{-1}$ have been ignored as the limit
$N\rightarrow\infty$ has been taken.  Returning to \erf{c}
gives the same QLE as for simple feedback.  In this rather
impractical way, all-optical feedback can replicate electro-optic
feedback.

\begin{figure}
\includegraphics[width=0.45\textwidth]{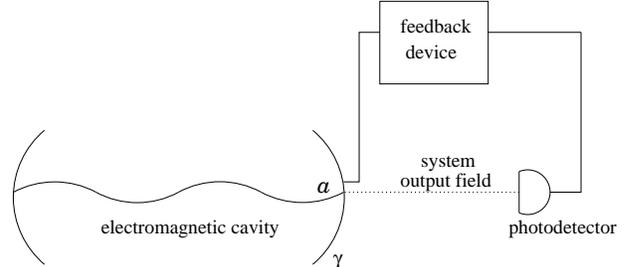}
\caption{\narrowtext Schematic representation of simple feedback.
The system is taken to
be a single mode optical cavity, with annihilation operator $a$ and
damping rate $\gamma$.  All further figures will also use an optical
cavity for the system.}
	\protect\label{sfb2}
\end{figure}
 \begin{figure}
\includegraphics[width=0.45\textwidth]{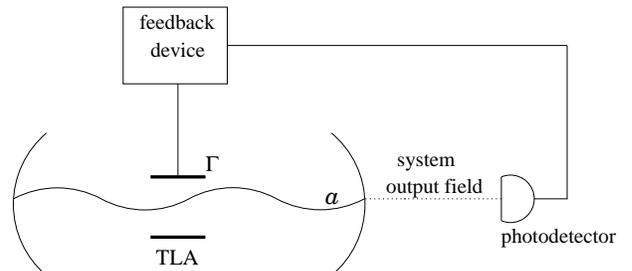}
\caption{\narrowtext Schematic representation of direct detection
feedback onto a TLA that is coupled back to the system.
The system damping rate has now been set equal to unity and the TLA
damping rate is $\Gamma$.}
	\protect\label{eoat2}
\end{figure}
 \begin{figure}
\includegraphics[width=0.45\textwidth]{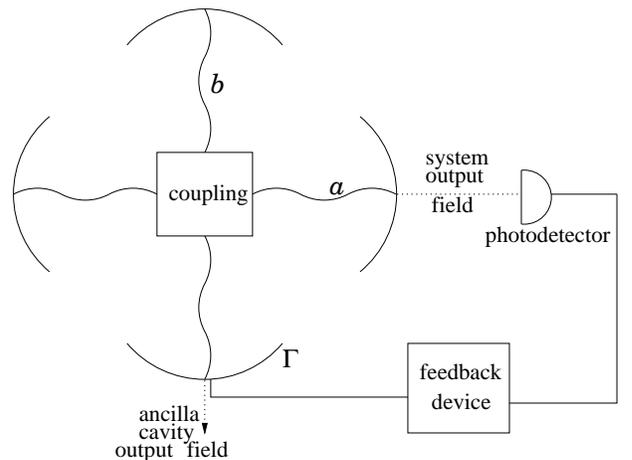}
\caption{\narrowtext Schematic representation of direct detection
feedback onto an optical cavity that is coupled back to the system.
The ancilla cavity has annihilation operator $b$ and damping rate
$\Gamma$.}
	\protect\label{eocav2}
\end{figure}
 \begin{figure}
\includegraphics[width=0.45\textwidth]{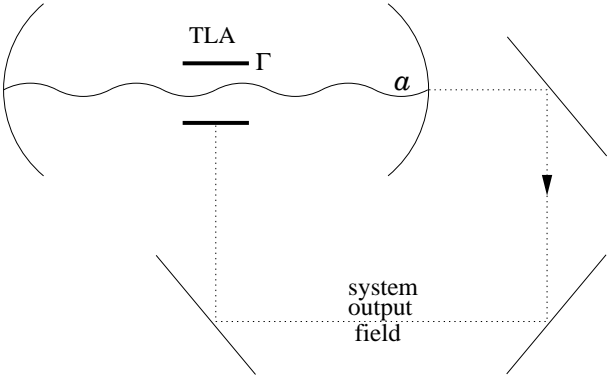}
\caption{\narrowtext Schematic representation of all-optical feedback
onto a TLA that is coupled back to the system.}
	\protect\label{alloptat2}
\end{figure}
 \begin{figure}
\includegraphics[width=0.45\textwidth]{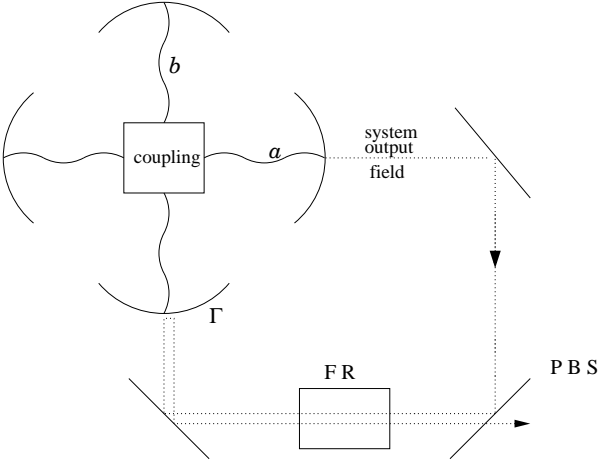}
\caption{\narrowtext Schematic representation of all-optical feedback
onto a single mode cavity that is coupled back to the system.  A
Faraday rotator (FR) and a Polarization dependent Beam Splitter (PBS)
are included in the feedback loop.}
	\protect\label{alloptcav2}
\end{figure}
  \begin{figure}
\includegraphics[width=0.45\textwidth]{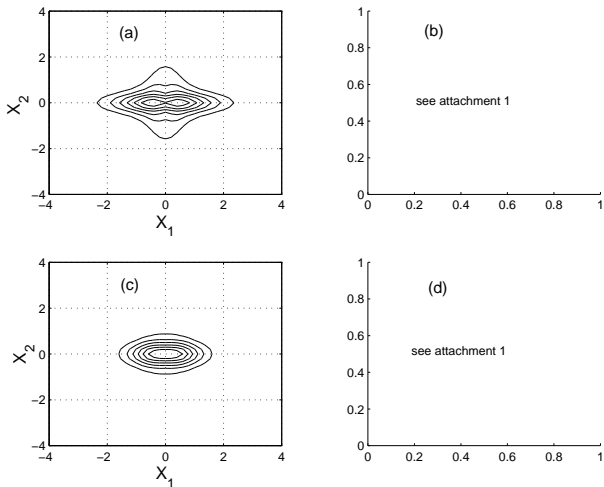}
\caption{\narrowtext Wigner functions of the steady states produced with
simple feedback for $\chi =\pi /2$.  Figs. (a) and (b) have $\lambda =2.2$,
while (c) and (d) have $\lambda =0.97$.  The mesh plots are included
to aid the readers interpretation of
the contour plots.}
	\protect\label{sfbsubplot}
\end{figure}
\begin{figure}
\includegraphics[width=0.45\textwidth]{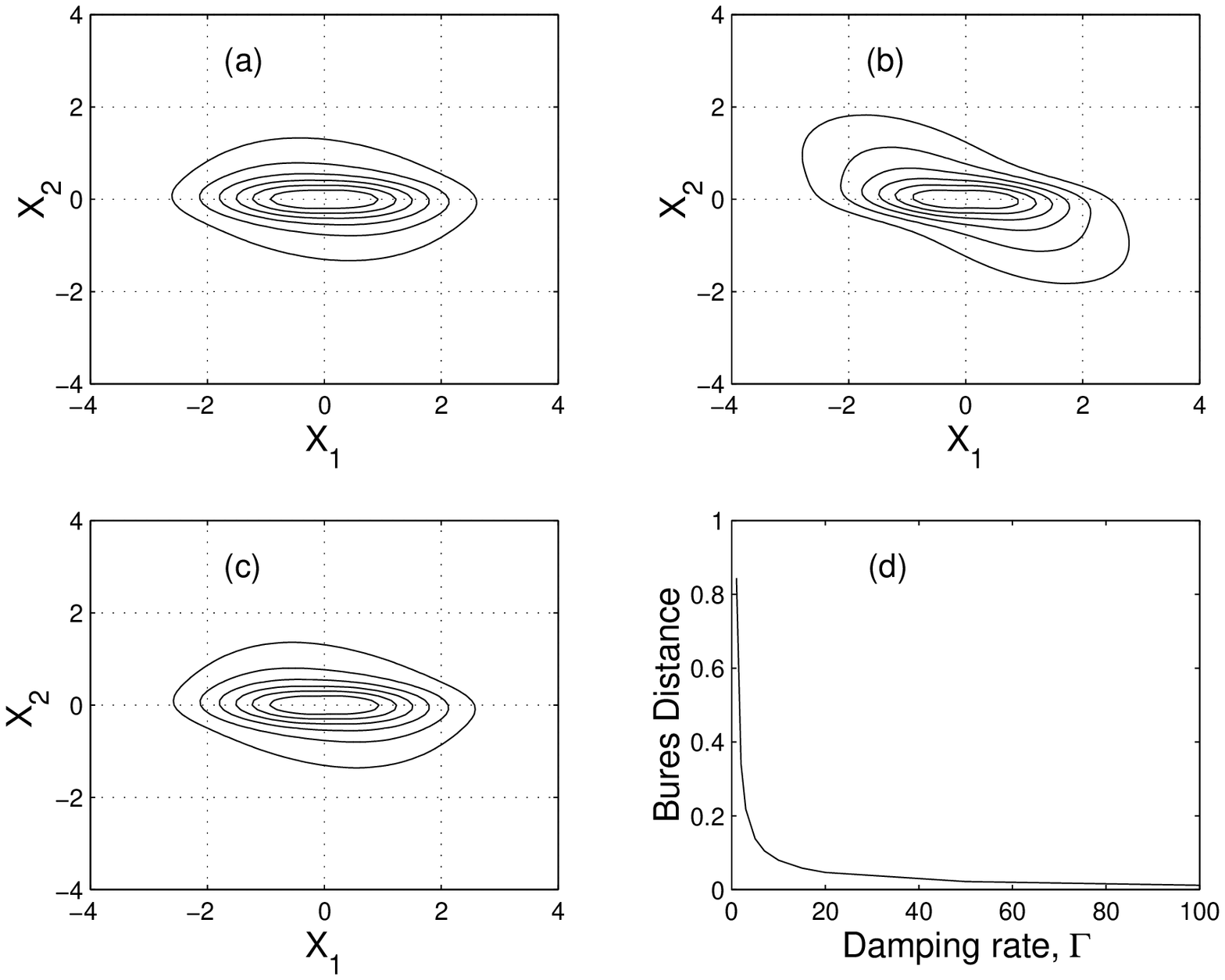}
\caption{\narrowtext Figs. (a), (b) and (c) are
Wigner functions of the steady states produced with
electro-optic feedback onto a TLA for $\chi = \pi /2$ and $\lambda =2.2$.
Fig. (a) is the adiabatically eliminated state. Figs. (b) and (c)
represent the full dynamics with $\Gamma =2$ and $20$ respectively.
Fig. (d) shows the Bures distance between the adiabatically
eliminated state and the state produced with the full dynamics as
$\Gamma$ increases.}
	\protect\label{eoatsubplot}
\end{figure}
\begin{figure}
\includegraphics[width=0.45\textwidth]{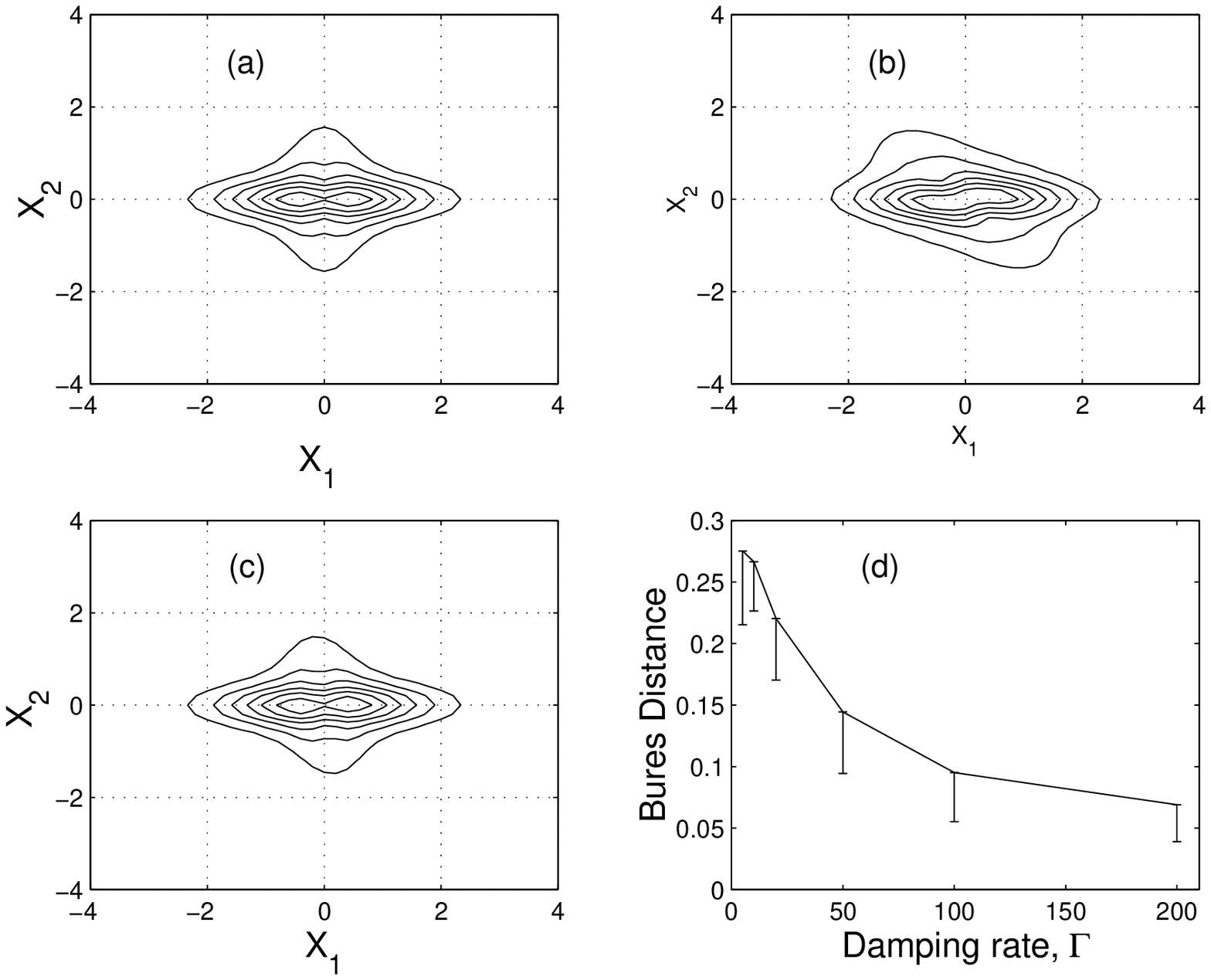}
\caption{\narrowtext Figs. (a), (b) and (c) are
Wigner functions of the steady states produced with
electro-optic feedback onto a mode for $\chi = \pi /2$ and $\lambda =2.2$.
Fig. (a) is the adiabatically eliminated state. Figs. (b) and (c)
represent the full dynamics with $\Gamma =10$ and $100$ respectively.
Fig. (d) shows the Bures distance between the adiabatically
eliminated state and the state produced with the full dynamics as
$\Gamma$ increases.  The error bars are due to
statistical error due to averaging over
a less than infinite number
of quantum trajectories.
Only half error bars are given because, in a high-dimensional Hilbert
space, a state with statistical errors
will tend to be further away from the adiabatically eliminated state
than the true ensemble average will be.}
	\protect\label{eocavsubplot}
\end{figure}
\begin{figure}
\includegraphics[width=0.45\textwidth]{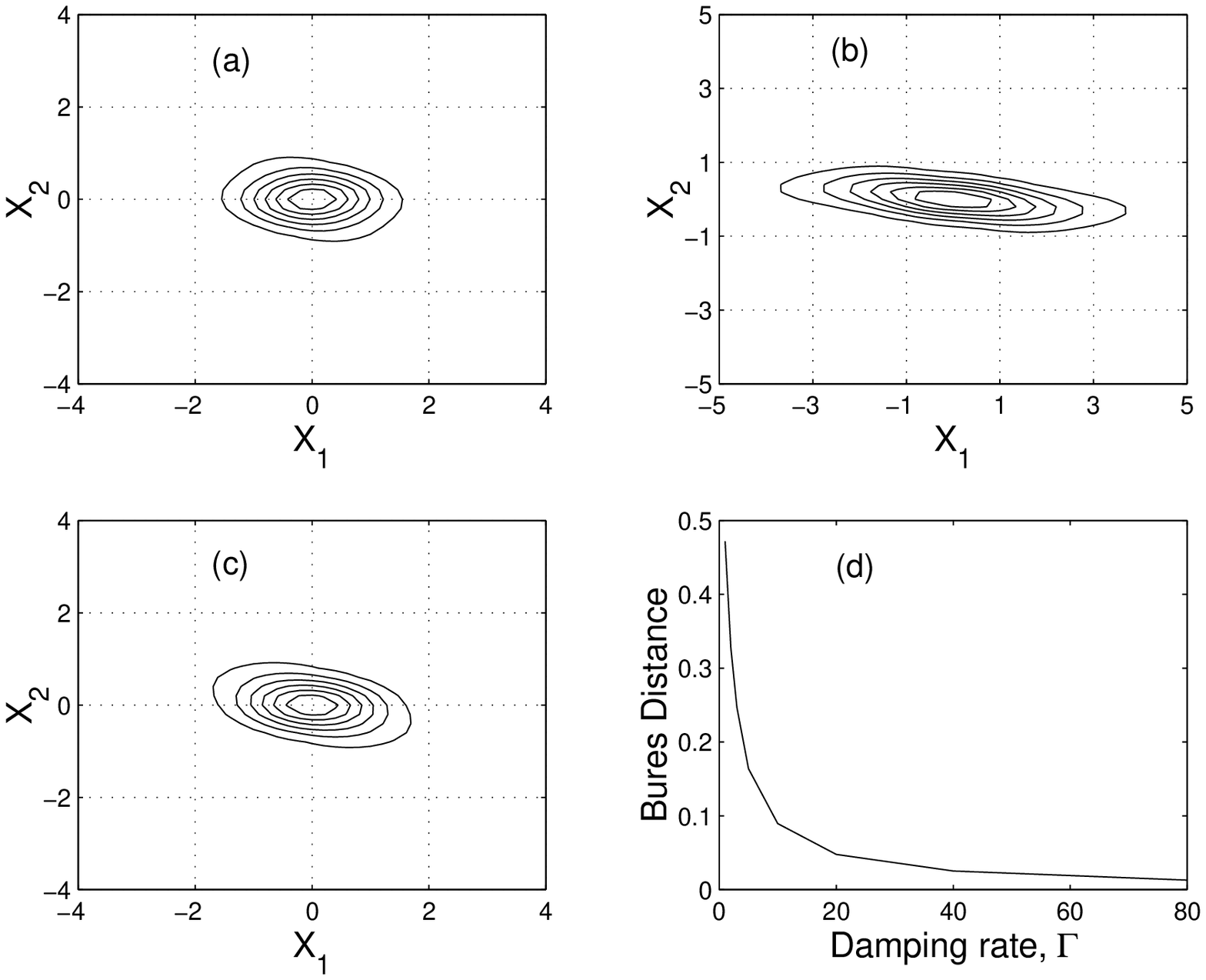}
\caption{\narrowtext Figs. (a), (b) and (c) are
Wigner functions of the steady states produced with
all-optical feedback onto a TLA for $\chi = \pi /2$ and $\lambda =0.97$.
Fig. (a) is the adiabatically eliminated state. Figs. (b) and (c)
represent the full dynamics with $\Gamma =1$ and $10$ respectively.
Fig. (d) shows the Bures distance between the adiabatically
eliminated state and the state produced with the full dynamics as
$\Gamma$ increases.}
	\protect\label{alloptatsubplot}
\end{figure}
\begin{figure}
\includegraphics[width=0.45\textwidth]{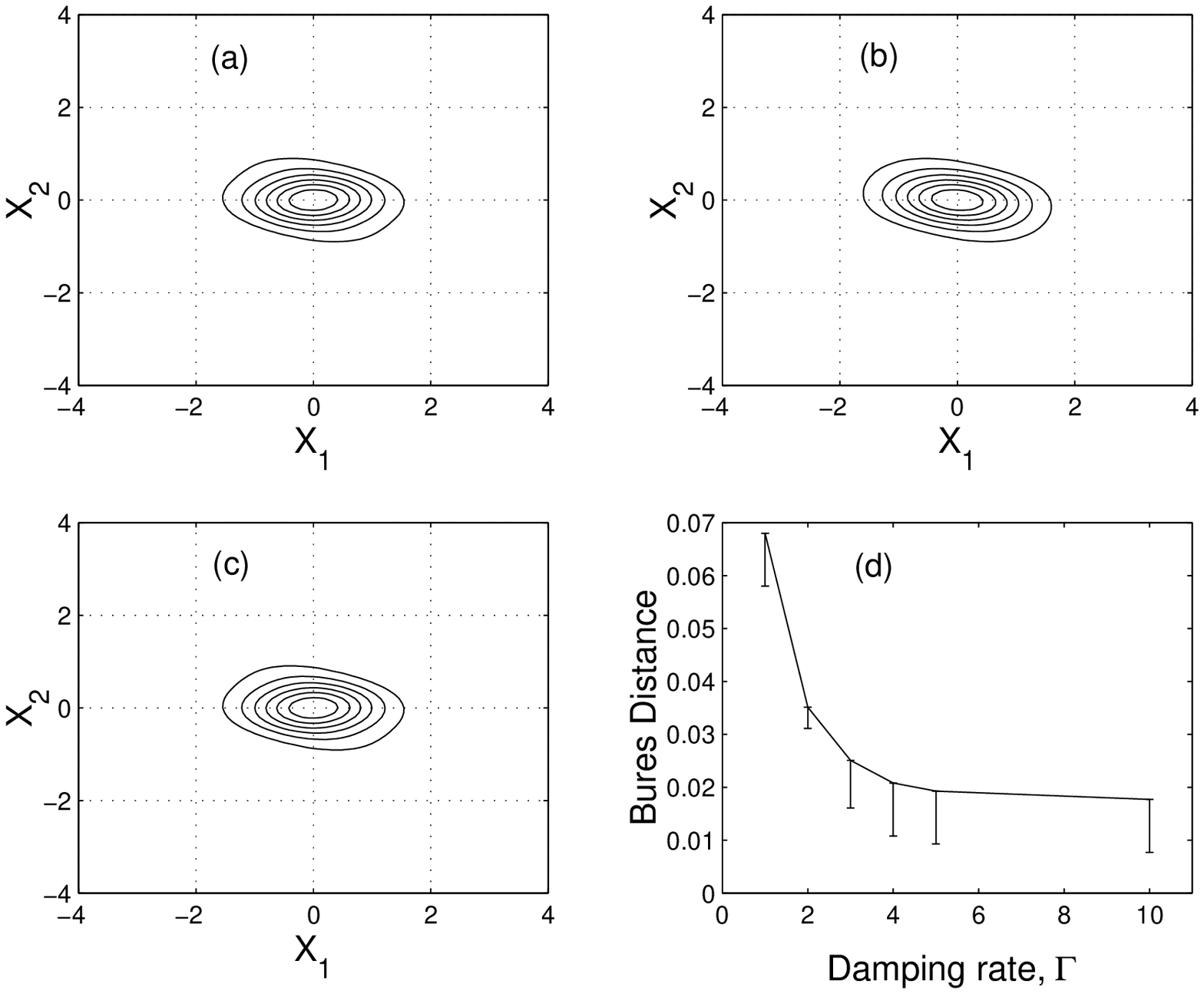}
\caption{\narrowtext Figs. (a), (b) and (c) are
Wigner functions of the steady states produced with
all-optical feedback onto a mode for $\chi = \pi /2$ and $\lambda =0.97$.
Fig. (a) is the adiabatically eliminated state. Figs. (b) and (c)
represent the full dynamics with $\Gamma =1$ and $10$ respectively.
Fig. (d) shows the Bures distance between the adiabatically
eliminated state and the state produced with the full dynamics as
$\Gamma$ increases.  Half error bars are used for the same reason as
in figure~\ref{eocavsubplot}.}
	\protect\label{alloptcavsubplot}
\end{figure}
\begin{figure}
\includegraphics[width=0.45\textwidth]{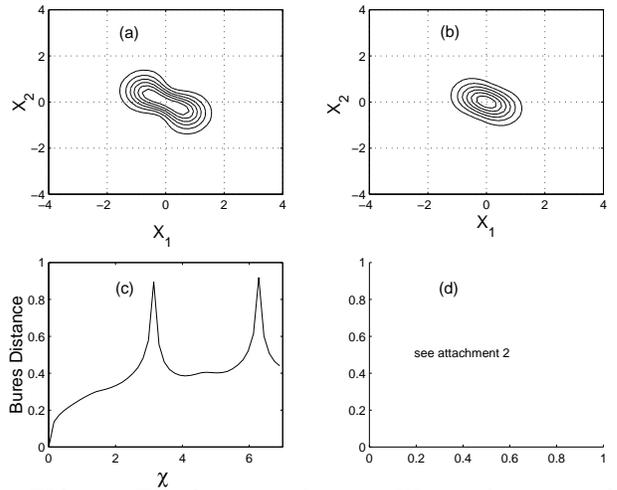}
\caption{\narrowtext The first two plots are
Wigner functions of the steady states produced with a $\chi^{(3)}=\pi /2$
non-linearity.  Figs. (a) and (b) have driving strengths of
$\lambda =2.2$ and $0.97$ respectively.  Fig. (c) shows the Bures
distance between simple and ``reversible'' feedback for $\lambda
=0.97$.  Fig. (d) is a mesh plot of the Wigner function displayed in
Fig. (a).}
	\protect\label{chisubplot}
\end{figure}
\begin{figure}
\includegraphics[width=0.45\textwidth]{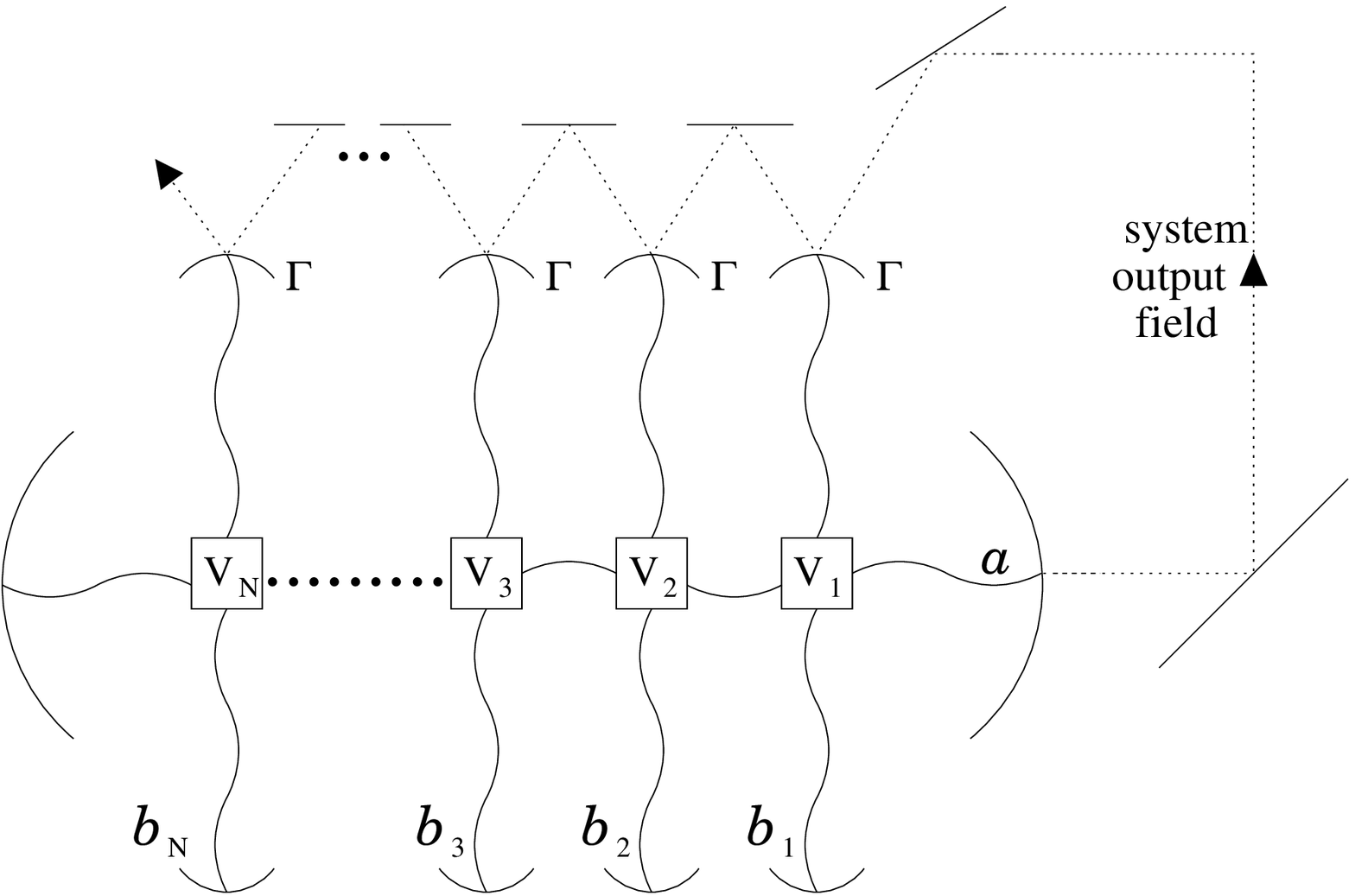}
\caption{\narrowtext Schematic representation of all-optical feedback
onto a very large number of optical cavities that are coupled back to
the system.  The $j^{{\rm th}}$ ancilla cavity has damping rate $\Gamma$
and annihilation operator $b_{j}$.}
	\protect\label{alloptinf2}
\end{figure}
 \end{multicols}

\begin{references}
\bibitem{fn0}
For example, the in-loop spectrum in Ref.~\cite{SloMil94}
is not correct because it does not take into
account the correction due to feedback as in Eq.~(2.62) of
Ref.~\cite{Wis94a}. Another paper \cite{SloMil95}
by the same authors also considering feedback in a compound system with
adiabatic elimination has worse problems: the central equation (9)
is not even dimensionally correct.
\bibitem{endnote}
An accurate simulation is taken to mean $\rho_{35,35}<1e^{-4}$.
\end{references}
\end{document}